\documentclass[amsmath,amssymb,aps,10pt,prd,letterpaper,nofootinbib,balancelastpage,notitlepage,superscriptaddress,twocolumn,floatfix]{revtex4-2}
\usepackage{graphicx}	
\usepackage{epstopdf}	
\usepackage{dcolumn}	
\usepackage{ragged2e}	
\usepackage{bm}		    
\usepackage{dsfont}     %
\usepackage[sort&compress]{natbib}	 	
	\setcitestyle{square,numbers,comma}	
\usepackage[colorlinks=true,urlcolor=blue,linkcolor=black,citecolor=red]{hyperref}	
\newcommand{\tabref}[2][]{Tab{#1}.~\ref{tab:#2}}		
\newcommand{\figref}[2][]{Fig{#1}.~\ref{fig:#2}}		
\newcommand{\secref}[2][]{Sec{#1}.~\ref{sec:#2}}		
\newcommand{\appref}[2][x]{Appendi{#1}~\ref{app:#2}}	
\renewcommand{\eqref}[2][]{Eq{#1}.~(\ref{eq:#2})}		
\newcommand{\citeR}[2][]{Ref{#1}.~\cite{#2}}			
\newcommand{\orcid}[1]{\href{https://orcid.org/#1}{\,\includegraphics[width=8px]{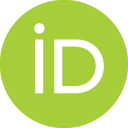}}}
\newcommand{\lb}{\ensuremath{\left}}						
\newcommand{\rb}{\ensuremath{\right}}
\newcommand{\ra}{\ensuremath{\rightarrow}}					
\newcommand{\Ra}{\ensuremath{\Rightarrow}}					
\newcommand{\nl}{\nonumber \\ & \quad }						

\begin{document}

\title{Gravity Gradient Noise from Asteroids}
\date{\today}

\author{Michael A.~Fedderke\orcid{0000-0002-1319-1622}}
\email{mfedderke@jhu.edu}
\affiliation{Department of Physics and Astronomy, The Johns Hopkins University, Baltimore, MD  21218, USA}
\affiliation{Department of Physics, Stanford University, Stanford, CA 94305, USA}
\author{Peter W.~Graham\orcid{0000-0002-1600-1601}}
\email{pwgraham@stanford.edu}
\affiliation{Department of Physics, Stanford University, Stanford, CA 94305, USA}
\author{Surjeet Rajendran\orcid{0000-0001-9915-3573}\,}
\email{srajend4@jhu.edu}
\affiliation{Department of Physics and Astronomy, The Johns Hopkins University, Baltimore, MD  21218, USA}

\begin{abstract}
The gravitational coupling of nearby massive bodies to test masses in a gravitational wave (GW) detector cannot be shielded, and gives rise to `gravity gradient noise' (GGN) in the detector.
In this paper we show that for any GW detector using local test masses in the Inner Solar System, the GGN from the motion of the field of $\sim 10^5$ Inner Solar System asteroids presents an irreducible noise floor for the detection of GW that rises exponentially at low frequencies.
This severely limits prospects for GW detection using local test masses for frequencies $f_{\textsc{gw}} \lesssim (\text{few})\times 10^{-7}$\,Hz.
At higher frequencies, we find that the asteroid GGN falls rapidly enough that detection may be possible; however, the incompleteness of existing asteroid catalogs with regard to small bodies makes this an open question around $f_{\textsc{gw}}\sim \mu$Hz, and further study is warranted.
We show that a detector network placed in the Outer Solar System would not be overwhelmed by this noise above $\sim 10\,$nHz, and make comments on alternative approaches that could overcome the limitations of local test masses for GW detection in the $\sim 10\,$nHz--$\mu$Hz band.
\end{abstract}
\maketitle

\tableofcontents

\section{Introduction}
\label{sec:intro}
The historic discovery of gravitational waves from multiple types of compact object mergers by the LIGO/Virgo collaborations \cite{PhysRevLett.116.061102,PhysRevLett.119.161101,PhysRevX.9.031040,Abbott:2020niy} has opened a new window into the Universe.
These instruments, along with the newly commissioned KAGRA \cite{Akutsu:2020zlw}, are sensitive to gravitational waves in the frequency band above 10\,Hz.
Much like in electromagnetism, there is an exceptionally strong case to probe other parts of the gravitational wave (GW) spectrum, since it is likely that the Universe has interesting secrets across the whole spectrum.
This case is particularly robust at lower frequencies where there are a number of expected astrophysical sources, such as white dwarf and neutron star binaries, as well as merging massive and supermassive black holes (see, e.g., \citeR[s]{Sesana:2019vho,Baibhav:2019rsa,Sedda:2019uro,Baker:2019pnp}).
A number of experiments are currently underway or proposed to investigate various parts of this spectrum.
In addition to LIGO/Virgo/KAGRA, this includes pulsar timing arrays (e.g., EPTA/LEAP, PPTA, NANOGrav, IPTA) that operate around 1--10\,nHz \cite{Kramer_2013,Kerr_2020,Arzoumanian:2020vkk,2016MNRAS.458.1267V}; the LISA constellation \cite{Baker:2019nia,LISA_Sci_Req,amaroseoane2017laser,PhysRevLett.120.061101} around 1--100\,mHz and TianQin~\cite{Luo:2015ght,Milyukov:2020fm} around 0.01--1\,Hz; `mid-band' proposals such as MAGIS/MIGA/AION \cite{Dimopoulos:2007cj,Dimopoulos:2008sv,Hogan:2010fz,Graham:2017pmn,Coleman:2018ozp,Canuel:2018fq,Badurina:2019hst}, and SAGE~\cite{Tino:2019tkb} around 1\,Hz; the Big Bang Observer (BBO) \cite{BBO,Crowder:2005nr}, which would bridge the LISA band and mid-band; the DECIGO mission \cite{Seto:2001qf,Kawamura:2018esd,kawamura2020current} in the 0.1--10\,Hz range; and the Einstein Telescope (ET) around 10--$\text{(few}\times10^2)$\,Hz \cite{Maggiore:2019uih}.

At present, there are no well-established techniques to probe gravitational waves in the 10\,nHz--0.1\,mHz band.
A preliminary investigation was performed in \citeR{Sesana:2019vho}, leading to the $\mu$Ares mission concept.
The science case discussed therein estimated that characteristic strain sensitivities $\sim 10^{-16}$--$10^{-19}$ were necessary in this band (at 100\,nHz--0.1\,mHz, respectively). 
It is an open question as to what the optimal technology is to achieve this strain sensitivity.

A gravitational wave detector fundamentally measures the disturbance in the space-time between two inertial test masses.
These test masses could either be distant astrophysical objects such as the neutron stars used in pulsar timing arrays, or environmentally isolated local proof masses as is the case in all the other experiments.
What kind of proof mass is necessary to probe 10\,nHz--0.1\,mHz gravitational waves?
The answer to this question is determined in part by the fundamental sources of noise that such a proof mass has to overcome. 

Gravity gradient noise (GGN),%
\footnote{\label{ftnt:akaGGN}%
    Also known as `Newtonian noise' \cite{Harms:2015awd}.
    }
which arises due to the gravitational force exerted by moving bodies on a test mass, is an irreducible form of noise in gravitational wave detection.
This noise cannot be shielded and it forces terrestrial gravitational wave detectors that use local proof masses to operate above $\sim 1$\,Hz \cite{Harms:2015awd,PhysRevD.86.102001,Coughlin_2016,PhysRevLett.121.221104,Buikema_2020}.%
\footnote{\label{ftnt:seismic}%
    Actual low-frequency sensitivity is in practice more typically limited by other noise sources; see, e.g., \citeR{Buikema_2020}.
    GGN does however constitute a fundamental limitation for ground-based experiments.
    } %
In this paper, we show that in the frequency band below $\sim (\text{few})\times 10^{-7}$\,Hz, detectors that use local proof masses (e.g., in a LISA-style constellation) and that are located within the Inner Solar System will be swamped by GGN arising from the motion of asteroids in the asteroid belt.

This conclusion is robust: it can be calculated using the measured properties of large asteroids in the belt.
As we will see, there are a significant ($\gg 50$) number of asteroids that contribute above the required noise floor.
This number is too large to permit the identification, resolution, and removal of every asteroid that contributes to this noise.
Consequently, gravitational wave detection in this band either requires the placement of detectors well into the outer parts of the Solar System, multi-constellation correlation techniques, or the use of distant astrophysical objects (which are not subject to this source of asteroid-induced GGN) as proof masses. 

The GGN from the asteroid belt drops rapidly at higher frequencies and is negligible above~$\sim \mu$Hz.
In this case, the noise is dominated by close encounters (well within 0.5\,AU) of asteroids with the proof masses.
We estimate this noise from the properties of detected and measured asteroids, and find it to be just small enough to permit detection.
However, this part of our detailed analysis is incomplete, since it is possible that a significant contribution to this noise could arise from asteroids that are either thus-far undetected, or whose properties are not presently well measured.
While this contribution is known to be sufficiently small at frequencies of interest to LISA \cite{LISA-Pre-Phase-A}, its effects around $\mu$Hz are less clear.
We defer detailed analysis of this point to future work, but we make some comments on the possible size of the effect, and highlight the need for the resolution of this point in order to establish the viability of using local test masses to search for gravitational waves around $\mu$Hz. 

The rest of this paper is organized as follows. 
In \secref{estimate}, we present an analytic estimate of the size and spectral features of the asteroid GGN.
In \secref{simulation}, we show the results of a numerical simulation that uses the measured properties of detected asteroids from the JPL Small-Body Database (JPL-SBD) \cite{JPL-SBD} to calculate the gravity gradient noise on a single-baseline LISA-style constellation located in the Inner Solar System.
We also give an estimate in \secref{closePasses} for the contribution from unmodeled close passes by asteroids that are either in the JPL-SBD but have incomplete physical information, or are not captured in the JPL-SBD.
We conclude in section \secref{conclusions}.  
A number of appendices give additional details: in \appref{diffAccn} we discuss technical aspects of our simulation, including the assumed detector orbits, the asteroid-induced accelerations on the detectors, and the asteroid orbits.
Also in \appref{diffAccn} and continuing in \appref{multipole}, we give a longer technical discussion of the analytic estimate presented in \secref{estimate}.
\appref{DFT} gives our conventions for the Discrete Fourier Transform (DFT), while \appref{windowing} gives a brief pedagogical introduction to the concept of windowing (apodization) of signals and its applicability to the computations presented in this work.

\section{Analytic Estimate}
\label{sec:estimate}

The qualitative features of the GGN expected from the asteroid belt can be understood by considering the following simplified model.
Assume that we have $N$ asteroids, with each asteroid in a circular heliocentric orbit that lies in the plane of the ecliptic.
Let $M_i$ and $R_i$ denote the mass and radius, respectively, of the $i^{\text{th}}$ asteroid ($i=1,\hdots,N$).
This asteroid orbits the Sun with a frequency $\omega_i = (G M_{\odot}/R_i^3 )^{1/2}$.
Consider a LISA-style GW detector consisting of a single baseline spanning between two orbiting proof masses that, when unperturbed, are separated by a fixed angle around a common heliocentric orbit in the plane of the ecliptic.
Let the orbital radius of the proof masses be $r < \min_i[ R_i]$ so that the baseline is contained fully within the asteroid belt, and let the unperturbed baseline length be $L \lesssim r$; see \appref{diffAccn} for further details.
The orbital frequency $\Omega$ of the proof masses in the constellation around the Sun is thus greater than $\omega_i$; let $\varpi_i \equiv \Omega-\omega_i$ be the relative asteroid--detector orbital frequency.

We consider this simplified model in detail in \appref[ces]{diffAccn} and \ref{app:multipole}, where we arrive at an algebraically complicated expression for the differential acceleration across the baseline due to the $i^{\text{th}}$ asteroid,  $\Delta a_i$, that is shown at \eqref{diffAccnFull} and which is expanded in various limits in \appref{multipole}. 
The reader interested in fine details should thus refer to those Appendices.
For the purposes of the present discussion however, we only intend to highlight certain qualitative features of these results, and supply an intuitive argument for their origin.
As such, it is useful to consider an approximate expression for $\Delta a_i$ which elides some detailed geometric baseline-projection and orientation effects.
$\Delta a_i$ can easily be understood to take the approximate form of a leading-order tidal acceleration: $\Delta a_i \sim GM_i L /  [\mathcal{R}_i(t)]^3$, where $\mathcal{R}_i(t)$ is the distance from the asteroid to the middle of the detector baseline, which varies as a function of time owing to the orbital kinematics of the asteroid and baseline: $\mathcal{R}^2_i(t) \sim R_i^2 + r^2 + 2rR_i \cos\lb( \varpi_i t + \alpha_i\rb)$, where $\alpha_i$ is the angular phase offset of asteroid $i$ around its circular orbit at time $t=0$. 
That is, we consider here the following approximate schematic form for $\Delta a_i$:%
\footnote{\label{ftnt:schematically}%
    We emphasize that the baseline-projected differential acceleration is actually the difference of two vector acceleration terms, projected onto the rotating baseline vector, and takes a form somewhat more algebraically complicated than that shown in \eqref{diffAccn}; see \appref{diffAccn} and \eqref{diffAccnFull}.
    Most of the qualitative features of the full result that are important for the argument here are however captured schematically by \eqref{diffAccn}; see \appref{multipole}, and \eqref{extraFactor} and the surrounding discussion.
} %
\begin{align}
    \Delta a_i &\sim \frac{G M_i L}{\big[R_i^2 + r^2 - 2rR_i \cos \lb(\varpi_i t + \alpha_i \rb)\big]^{3/2}} \label{eq:diffAccn} \\
    & = \frac{G M_i L}{R_i^3}\left[1 +  \sum_{j = 1}^{\infty} c_j \left(\frac{r}{R_i}\right)^j \cos^j\left( \varpi_i t +\alpha_i\right)\right]
    \label{eq:diffAccnExp}
\end{align}
where the $c_j$ are numerical coefficients that can be computed from a multipole expansion of \eqref{diffAccn}.%
\footnote{\label{ftnt:moreComplicatedFull}%
    A multipole expansion of the full result \eqref{diffAccnFull} yields the significantly more algebraically complicated result at \eqref{inside} but, again, \eqref{diffAccnExp} schematically captures the qualitative features relevant to the argument here; see \appref{multipole}.
    } %

The crucial property of the expansion at \eqref{diffAccnExp} that is relevant is that the Fourier amplitude of the harmonic $\omega_q \equiv q \varpi_i$ is suppressed rapidly%
\footnote{\label{ftnt:expansionOfCos}%
    This observation follows immediately from \eqref{diffAccnExp} by noting that a term $\sim\cos\lb[ q \lb( \varpi_i t + \alpha_i \rb)\rb]$ occurs in the expansion of $\cos^k\lb( \varpi_i t + \alpha_i \rb)$ for $k=q$ (and for all $k>q$ that have the same parity as $q$), but does not appear for any $k<q$.
    This conclusion holds, with some minor modifications, for the expansion of the full result \eqref{diffAccnFull}; see \appref{multipole}.
    } %
as $\left(r/R_i\right)^q\ll1$ when $q \gg 1$ (i.e., at high frequencies), while we expect unsuppressed Fourier power at frequencies in the vicinity%
\footnote{\label{ftnt:detailedDifferences}%
    For $r<R_i$, the full results in \appref{multipoleInside} show that the peak acceleration Fourier power is actually at $\omega\approx 2\varpi_i$; while we note this for completeness, this detail is not relevant at the level of the present discussion.
    } %
of $\omega \sim \varpi_i$.
For a detector in the Inner Solar System, we thus expect the noise from the asteroid belt to be most important in the vicinity of the corresponding orbital frequencies $\sim 10$--$100$\,nHz, becoming smaller rapidly at higher frequencies. 

In the $\sim 10$--$100$\,nHz band, the limit on the detectable characteristic strain $h_c$ of a gravitational wave imposed by this asteroid GGN can be estimated as follows.
Suppose there are $N_i$ asteroids of mass $M_i$, and that they are located at random phase offsets $\alpha_i$ around the circular orbit, with uniform probability.
Since the bulk of the belt asteroids are all roughly at the same radius $R_i\sim R$ from the Sun [within $\mathcal{O}(1)$ factors], the total acceleration noise in this frequency band from these $N_i$ asteroids is, for a typical random configuration, $\Delta a \sim ( G M_i L/R^3 )\cdot \sqrt{N_i}$, where this scaling with asteroid number arises owing to the randomness of the asteroid locations around the orbit; see also \appref{multipole}.%
\footnote{\label{ftnt:randomAsteroids}%
    More precisely, this is the parametric scaling of the ensemble average over asteroid configurations of the amplitude of $\Delta a$ at the dominant frequency; however, when viewed as a function of the random asteroid configuration, $\Delta a$ is a random variable which is distributed proportional to a $\chi_2$ distribution.
    It thus has a variance of the same order as the mean; in any one random asteroid configuration then, the result for the amplitude of the acceleration at the dominant frequencies could be smaller or larger by a factor that is generically, although not necessarily, $\mathcal{O}(1)$.
    We also note that in the unphysical case where the asteroids are exactly periodically spaced around their orbit, the scaling with $N_i$ is faster: $\Delta a \propto N_i$; however, this scaling is only realized at certain specific frequencies in such cases, owing to phase cancellations.
    } %

Comparing this noise to the differential acceleration $\Delta a_{\textsc{gw}} \sim h_c L \omega^2$ produced by the gravitational wave, we get a strain $h_c \sim ( G  M_i \sqrt{N_i} ) / ( \omega^2 R^3 )$.
The data from the JPL-SBD indicate that the number of asteroids with a diameter $d \gtrsim \text{km}$ is roughly approximated as $N(d) \sim  (1000\,\text{km}/d)^2$.
Since the mass of an asteroid is $M_i \propto d_i^3$, the magnitude of this acceleration noise is dominated by the heaviest asteroids.
But this magnitude does not set the noise floor since there are only a small number of heavy asteroids and thus it should be possible to independently resolve and remove them from the data-stream.
The limit is instead set by the smallest value of $N_i$ that is too large to be individually resolvable; i.e., the highest-mass asteroids that are also sufficiently numerous so as to be unresolvable.
In a gravitational wave mission operating for $\sim$ 10 years in the frequency band 10--100~nHz, it will be hard to resolve more than $\sim 10$ asteroids.%
\footnote{\label{ftnt:optical}%
    This is on the basis of the acceleration data alone.
    It is possible that with independent electromagnetic-spectrum (optical, radar, etc.) observations of asteroid locations, more of these objects' waveforms could be removed from the data. 
    However, in order to avoid floating a large number of free asteroid mass parameters in some sort of fit to remove those objects from the acceleration waveform (which could lead to possible degeneracies between a sum of asteroid waveforms of unknown normalization and a real gravitational wave signal), precise and accurate independent asteroid mass determinations would be required.
    Unfortunately, such independent asteroid mass determinations are difficult to obtain and are only available for a small number of asteroids, limiting the utility of such an approach.
    } %
Conservatively taking this number to be $N_i \sim 50$, we find that asteroids with a diameter $d\gtrsim 170$\,km will be the dominant source of this noise.
Assuming a mean asteroid mass-density of $\rho \sim 3\,\text{g}/\text{cm}^{3}$ and placing the asteroid belt at $R \sim 3$\,AU, this yields a limiting characteristic strain $h_c \sim  10^{-12}$.
This is about 4 orders of magnitude larger than minimum strain sensitivity estimated to be necessary to see interesting gravitational wave sources in this band \cite{Sesana:2019vho}. 

This noise is clearly a limiting factor at 100\,nHz, while it will fall rapidly at higher frequencies as long as the detector constellation is well within (or, indeed, well outside) the asteroid belt.
To map the precise limit across all these frequencies, taking into account a realistic asteroid belt population, it is necessary to perform a numerical simulation based on the objects contained in the JPL-SBD; such a simulation is presented and discussed in the next section.

While the asteroids in the asteroid belt cause suppressed noise at frequencies above $\sim\mu$Hz, a detector also experiences accelerations from closer encounters with asteroids.
If these close passes are sufficiently numerous as to be unresolvable, this will be an additional source of noise.
We also consider these kinds of encounters in the simulation discussed below using the objects that are cataloged in the JPL-SBD, but our results are incomplete as the JPL-SBD is not an exhaustive catalog for objects of this class (owing either to missing physical data, or to catalog incompleteness). 
Although we defer detailed consideration of this point to future work, we supply a rough argument as to the possible size of the noise generated by such unmodeled asteroids in \secref{closePasses}.

\section{Simulation and Results}
\label{sec:simulation}

While the simplified model discussed in \secref{estimate} (and developed in more detail in \appref{multipole}) yields a useful qualitative understanding of the frequency content of the GGN from a belt-like population of asteroids, obtaining a quantitative estimate of the actual asteroid GGN from the real asteroid belt population requires numerical simulation.
We detail our approach to such a numerical simulation in this section.

A numerical approach is required because the simplified model so far discussed does not consider the effects of elliptical,%
\footnote{\label{ftnt:elliptical}%
    Indeed, analytical understanding of the frequency content of the acceleration induced by even a single asteroid on an orbit with non-zero eccentricity is intractable because the temporal evolution of the radius and angular displacement of a body in an elliptical orbit under the action of (even Newtonian) gravity cannot be expressed in closed form in terms of elementary functions amenable to analytical Fourier analysis.
    } %
or out-of-the-plane-of-the-ecliptic, asteroid orbits; additionally, we must take into account the realistic distribution, and the correlations between, the physical (diameter and mass) and orbital (semimajor axis, eccentricity, orbital orientation, and time of perihelion passage) parameters of the asteroids in the belt.

\subsection{Simulation description}
\label{sec:simulationDesc}

At a high level, the simulation we perform is straightforward to describe. 
We begin by setting up a detector network of two test masses that are assumed to be on exactly circular heliocentric orbits that lie exactly in the plane of the ecliptic of the Solar System (i.e., orbital inclination $\hat{\iota}_{\text{det.}}=0$), with a fixed baseline distance $L$ separating them; see \appref{detectors}.
We assume throughout the simulation that perturbations to the locations of the detectors are small enough to be neglected in computing accelerations on the test masses due to asteroids (i.e., all accelerations on the test masses are computed assuming the \emph{unperturbed} positions of the test masses).

For each asteroid or comet (hereinafter, `object') $i$ in the JPL-SBD for which a diameter measurement $d_i$ is reported, we estimate the asteroid mass%
\footnote{\label{ftnt:masses}%
    Direct mass measurements (or, more specifically, the product $GM_i$) are only available for a very small number of objects.
} %
as $M_i = (\pi/6) \rho d_i^3 $, assuming (1) that even though most objects are non-spherical, the JPL-SBD diameters%
\footnote{\label{ftnt:diameters}%
    The diameters of most objects in the JPL-SBD are not directly measured, but are inferred either from the absolute visual magnitude and geometric albedo of the object \cite{JPL-SBD-phys}, or using other physical data (see, e.g., \citeR{2014ApJ...791..121M} for one example).
    Some of the larger objects do however have directly measured diameters \cite{JPL-SBD-phys}.
} %
can be used in this way to obtain an acceptably accurate approximation to the volume of the object, and (2) that all objects have the same uniform mass-density of $\rho = 2.5\,\text{g/cm}^3$, which is approximately the density of large asteroids such as Ceres for which direct mass and volume measurements are available.%
\footnote{\label{ftnt:density}%
    Although we acknowledge that the densities of individual objects may be quite different, this is the best one can do absent additional data.
    } %

We then place object $i$ on an exactly elliptical heliocentric orbit using the (osculating) orbital parameters (semimajor axis, ellipticity, orbital orientation angles, and time of perihelion passage) that are reported in the JPL-SBD; see \appref{elliptical}.
Our simulation ignores any perturbations to these orbits that might arise from other bodies; that is, we do not perform a computationally expensive $N$-body simulation using the JPL-SBD objects, but instead assume that the dynamics of each object is governed by (Newtonian) two-body Sun--asteroid gravitation.%
\footnote{\label{ftnt:notNBody}%
    We of course acknowledge that this is known to yield inaccurate ephemerides (i.e., the exact asteroid position in the orbit at any given time) when considered over long time periods (with the errors growing in time as one moves away from the epoch at which the two-body elliptical orbital parameters were determined), even if the orbital shape and orientation are not strongly perturbed.
    Nevertheless, given our other approximations, we feel that the result summed over all asteroids will still yield an acceptable estimate for the differential acceleration on the test masses.} %

Once the position and mass of object $i$ are thus determined, we can then straightforwardly compute the contribution of object $i$ to the component of the differential acceleration on the detectors that lies along the instantaneous detector baseline direction, which we denote by $\Delta a_i$, and refer to as the `baseline-projected differential acceleration'; see \appref{generalOrbits} for the exact definitions and further technical details.
The simulation is performed so as to reflect a mission of total duration $T = 10\,\text{years}$ starting (somewhat arbitrarily) from the epoch of most recent Earth perihelion passage.%
\footnote{\label{ftnt:perihelion}%
   Specifically, we select the time $t_0$ defined at footnote \ref{ftnt:offset} to be within a few minutes of 8AM UT on January 5, 2020 \cite{JPL-HORIZONS}.} %
Within this duration, $\Delta a_i$ is sampled at some large number ($\mathcal{N}_i = 3000$) of discrete points that are taken to be equally spaced in time by $\Delta t = T/\mathcal{N}_i$, yielding the discrete time-series of baseline-projected differential acceleration contributions $\Delta a_i(t_j = j\Delta t)$.%
\footnote{\label{ftnt:DFT}%
    These parameters are sufficient to access the frequency range of interest: we remind the reader that in the analysis of discrete time-series, the discrete Fourier transform (DFT) yields results for the frequency content that are spaced by $\Delta f = 1/T$, and that are without aliasing effects up to the Nyquist frequency $f_{\text{Nyquist}} = 1/(2\Delta t) = \mathcal{N}_i/(2T) = (\mathcal{N}_i/2) \Delta f$. 
    To access the 10\,nHz--$\mu$Hz regime then, we minimally require $T \gtrsim 10^8\,\text{s} \sim 3\,\text{years}$, and $\mathcal{N}_i \gtrsim 10^2$.} %
We repeat this process for all $N = 140\,299$ objects in the JPL-SBD for which diameter data $d_i$ are available [such that a mass estimate $M_i(d_i)$ can be obtained]; while this represents only $\sim 15\%$ of the total of $\sim 10^6$ JPL-SBD objects, diameter data are not available for the remainder of these objects, so we cannot include them in the simulation (the mass of the object of course serves as the normalization of $\Delta a_i$).
The total baseline-projected differential acceleration on the detectors, $\Delta a(t_j)$, is then obtained by summing the contributions over all $N$ asteroids; see \appref{generalOrbits}.
Further data processing is discussed in the next few sections.

\subsubsection{Objects omitted from the simulation}
\label{sec:omittedObjects}
Before proceeding however, some further comments are in order regarding our choice to exclude the remaining $\sim85\%$ of objects in the JPL-SBD catalog that are without explicit diameter determinations $d_i$. 
While it is the case that (nearly) all of these objects do have an absolute magnitude $H_i$ supplied in the catalog, their (geometric) albedos $\alpha_i$ are not supplied.
Armed with both, we could have made an estimate of their diameter as $d_{H,i} \equiv d_H(H_i,\alpha_i) \sim \lb(1329\,\text{km}/\sqrt{\alpha_i}\rb)\exp[-0.2H_i]$; see, e.g., \citeR[s]{FOWLER_Asteroids,HARRIS1997450}.
While it is possible to replace $\alpha_i$ in this estimate with the average geometric albedo $\bar{\alpha} \approx 0.1$ of the objects in the JPL-SBD for which such data are available (i.e., the objects we simulated) in order to obtain a \emph{rough} idea of the diameters of these objects as $\bar{d}_{H,i} \equiv d_H(H_i,\bar{\alpha})$, the albedo distributions for asteroids are broad enough that $\bar{d}_{H,i}$ can easily be incorrect by a factor $>1$.%
\footnote{\label{ftnt:D_vs_DH}%
    For the simulated asteroids in the JPL-SBD for which an actual diameter $d_i$ as well as absolute magnitude $H_i$ and albedo $\alpha_i$ are all available, the estimate $d_{H,i}$ is highly correlated with $d_i$. 
    However, specific asteroids, particularly those in the $d_i \sim 1$--$100$\,km class, still exhibit differences between $d_i$ and $d_{H,i}$ of up to half an order of magnitude, although typically the difference is smaller.
    Additional systematic effects at the level of $\mathcal{O}(1)$ factors on the diameter determinations are also clear when comparing $\bar{d}_{H,i}$ to $d_i$ for this class.
    } %
Because the diameter of course enters raised to the third power in the mass estimate for any one asteroid, and because this class of asteroids is more numerous than the simulated class for which reliable diameter data are provided, following such an approach to include these objects in the simulation could give rise to large systematic effects in certain of our results.
We thus follow the conservative approach of simply omitting this class of objects; in the worst case, this would mean that our results would thus represent a conservative noise floor, which could be subject to upward revision if better diameter determinations for the class of objects that we do not simulate in this work become available. 
However, the extent to which we expect our results would be sensitive to corrections by these unmodeled objects is frequency dependent, and depends on the mass and orbital distributions of these objects.

It is thus important to roughly characterize the unsimulated population of JPL-SBD objects. 
The foregoing caveats about using $\bar{d}_{H,i}$ as a diameter estimate in the context of a detailed simulation notwithstanding, it does allow such a rough characterization.
We find that objects in the unmodeled class that are in the Inner Solar System (semimajor axis $a_i \lesssim 10\,$AU) typically have $\bar{d}_{H,i} \lesssim 30$\,km, with the vast majority of these objects having diameters of a $\bar{d}_{H,i} \sim \text{few}$ km.
The larger of these tend to have semimajor axes $a_i$ which are indicative of being part of the main asteroid belt (or further out), while the smaller of these tend to have semimajor axes which would indicate closer passage to the vicinity of a detector baseline at $r\sim 1\,$AU.
Studying the distributions in the $(a_i,\bar{d}_{H,i}$) plane of the unsimulated objects as compared to the distributions in the $(a_i,d_{i}$) plane of the objects we do simulate, we estimate that, as a general rule, the JPL-SBD objects that we have omitted from the simulation are smaller, and pass closer to the detectors, than those objects we have included in the simulation. 
Moreover, for objects in the Inner Solar System (semimajor axis $a_i \lesssim 10\,$AU), the simulated objects with $d_i\gtrsim 4$\,km are more numerous than the unsimulated objects with $\bar{d}_{H,i}\gtrsim 4$\,km, indicating that \emph{our simulation accurately captures the effect of all relevant asteroids with diameters larger than a few km}.
As our discussion later in this paper will make clear, the fact the simulated objects can thus be estimated to capture most of the larger objects that do not pass very close to the baseline will generally mean that our lower-frequency results should be completely insensitive to the absence of the unmodeled objects; on the other hand, the absence of the smaller, closer-passing objects implies that our higher frequency results may be more sensitive to them, and we give an estimate of this sensitivity in \secref{closePasses}.

We also note that there is a not insignificant population of trans-Neptunian objects with $\bar{d}_{H,i} \sim \text{few}\times 10^2\,$km that are also in the unsimulated class of objects.
However, since their semimajor axes $a_i \gtrsim 35\,$AU, their exclusion will not substantially impact our results. 
To estimate their contribution to $\Delta a$, note that there are $N^{\text{incl.}}_5\approx 650$ objects with diameters $d_i$ in the range 50--500km that we did simulate that have $a_i<5\,$AU, and we estimate that there are $N^{\text{excl.}}_{35}\approx 3200$ trans-Neptunian objects with $a_i>35$\,AU that have estimated diameters $\bar{d}_{H,i}$ in the same 50--500km range.%
\footnote{\label{ftnt:otherCount}%
	For completeness, note that there are only 4 objects with $50 \leq \bar{d}_{H,i} / \text{km} \leq 500$ and $a_i<5\,$AU that are in the unsimulated class; we capture more than 99\% of these larger asteroids that are in the belt in our simulation.
	} %
Because we expect contribution to $\Delta a \propto \sqrt{N}/R^3$ (see \secref{estimate}) and because all asteroids with $R \gg 1$\,AU contribute in roughly the same frequency range to the acceleration (at least for Inner Solar System detector orbits, since $\varpi_i \sim \omega_i$ in this case; see \secref{estimate}), we estimate that the impact of the omitted trans-Neptunian objects would be at the level of $\sim \sqrt{N^{\text{excl.}}_{35}/N^{\text{incl.}}_5}(5/35)^3 \sim 6\times 10^{-3}$ of that of the closer objects in the same diameter class that we did include.

Finally, note also that we separately completely neglect the effect of the planets (and Pluto) on the detectors, on the assumption that their positions and masses are known well enough to enable them to be efficiently removed during preliminary data processing.
For our purposes, this is also a conservative assumption, as the inability to remove such effects completely can only act to increase the GGN noise level.

\subsection{Fourier analysis and windowing}
\label{sec:windowing}
In order to obtain a baseline-projected differential acceleration noise spectrum that we can use to compare to the baseline-projected differential acceleration that would be induced by a gravitational wave in a matched filter search, the obvious next steps would be to simply take the discrete Fourier transform (DFT) of the time-series $\Delta a(t_n)$, and find the power spectral density (PSD) of the noise; see \appref{DFT} for our DFT conventions.

While we will ultimately proceed in this general fashion, we must first exercise some caution: it is well-known in signal processing that when applying the DFT to a dataset of duration $T$ that is the sum of multiple frequency components that have disparate power levels, the presence of lower-power components can be masked if there are higher-power components present that have a non-integer number of cycles within the analysis duration $T$,%
\footnote{\label{ftnt:altwording}%
	Another way to phrase this is that if the frequency $f$ of a signal component does not lie at exactly one of the discrete frequencies accessed by the DFT, $f_j = j \Delta f = j/T$ for $j=0,\ldots,\mathcal{N}-1$ (or, indeed, any higher frequency that differs from one of these DFT frequencies by an amount of exactly $\mathcal{N}/T$), then the number of cycles in the analysis duration, $N_{\text{cycles}} = fT \notin \mathbb{Z}$, is non-integer, and it is easy to show that a signal component at frequency $f$ will contribute power in every DFT frequency bin.} %
which is of course the generic case.
This is the phenomenon of `spectral leakage'; see \appref{windowing} for a pedagogical discussion.

Based on the simple analytical model of \secref{estimate} (see also \appref{multipole}), we expect to find ourselves in precisely this scenario: the Fourier power at higher harmonics of the fundamental frequency $\varpi_i$, which are still quite close to the fundamental, is expected to be (exponentially) suppressed compared to the Fourier power near the fundamental, at least when $r\nsim R_i$.
Therefore, in order to map out the true underlying noise power spectrum, and not just recover the leakage power spectrum of the dominant noise sources, we proceed by `windowing' (or `apodizing'), the asteroid-GGN-induced baseline-projected differential acceleration time-series $\Delta a(t)$ prior to taking the DFT; see \appref{windowing} or \citeR{Harris:1978wdg} for a pedagogical discussion.
In particular, we define
\begin{align}
    \Delta a_w(t) &\equiv w(t) \cdot \Delta a(t), \label{eq:windowedDeltaAccn}\\
    w(t) &\equiv \sin^8( \pi t/T ) \label{eq:windowFunc},
\end{align}
where $w(t)$ is the window function discussed in \appref{windowing} [see also \eqref{window}].
Working with the windowed $\Delta a_w(t)$ allows us to extract the baseline-projected differential acceleration spectrum of the asteroid GGN over a much larger dynamic range than would be possible for the unwindowed $\Delta a(t)$; the cost is that the result must be interpreted with some care in the comparison to a GW signal because the windowing would be applied also to a signal and would modify the normalization, as we discuss below in \secref{signal}.

Because windowing is a multiplication in the time domain, it is a convolution in the frequency domain [cf.~\eqref{convolution}]:
\begin{align}
    \widetilde{\Delta a}_w(f_k) &= \frac{1}{T} \sum_{n=0}^{\mathcal{N}-1} \widetilde{\Delta a}(f_n) \cdot \tilde{w}(f_{(k-n)\!\!\!\!\mod \mathcal{N}});\\
    &\phantom{=} k = 0,\hdots,\mathcal{N}-1,\nonumber
     \label{eq:hwDFT1}
\end{align}
where $f_k \equiv k \Delta f \equiv k/T$.

Therefore, we straightforwardly compute the DFT of the baseline-projected differential acceleration, convolve it with the DFT of the window function to obtain the DFT of the windowed baseline-projected differential acceleration, and then finally compute the one-sided PSD [see \eqref{PSD}] of the latter, $S_k^{\textsc{ggn}}[\Delta a_w] \equiv S^{\textsc{ggn}}[\Delta a_w](f_k)$, as the measure of the spectrum of the asteroid GGN.

\subsection{Results}
\label{sec:results}

\begin{figure}[p]
\includegraphics[width=0.9\columnwidth]{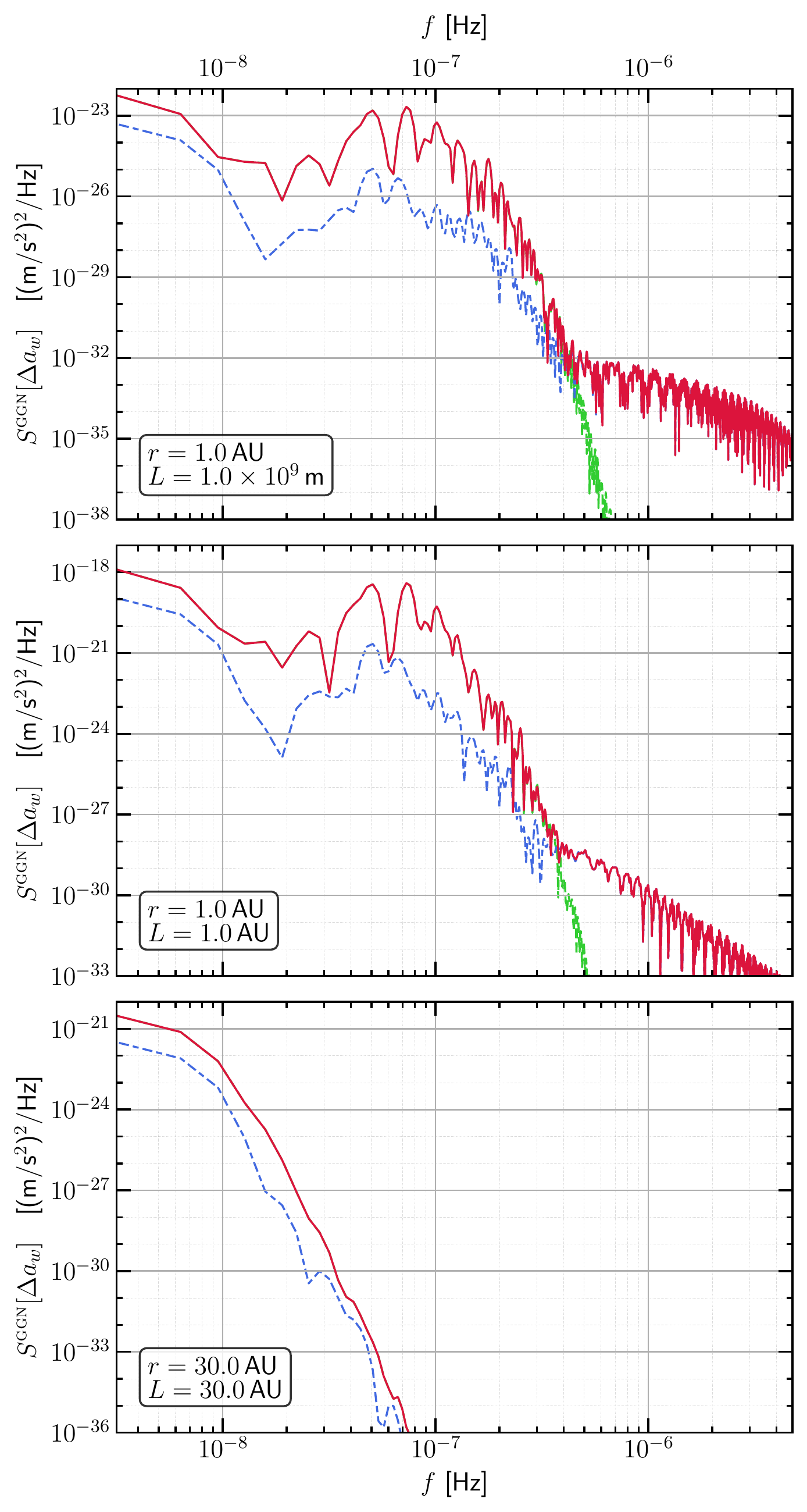}
\caption{\label{fig:accnResults}%
    Windowed baseline-projected differential acceleration PSD due to gravity gradient noise induced by asteroids and comets in the JPL-SBD on a pair of detectors at heliocentric orbital radius $r$ with a baseline $L$, as annotated.
    The solid red lines show the PSD $S^{\textsc{ggn}}[\Delta a_w]$ of $\Delta a_w$ as defined at \eqref[s]{windowedDeltaAccn} and (\ref{eq:DeltaADefn}), computed per the numerical simulation procedures outlined in \secref[s]{simulationDesc} and \ref{sec:windowing}.
    These results account only for those objects in the JPL-SBD with an explicitly supplied diameter (see \secref{simulationDesc} for discussion), and for the Inner Solar System baselines they may thus represent an underestimate of the noise at higher frequencies ($f \gtrsim \text{few}\times 10^{-7}\,$Hz); see the detailed discussion around \figref{higherClosePassNoise} in \secref{closePasses}.
    We assume a mission duration of $T = 10\,\text{years}$ and $\mathcal{N}=3000$ sampling points in the simulation, and results are displayed for frequencies $f \in [ \Delta f , f_{\text{Nyquist}} ]$.
    Note that the frequency $f$ shown here is simply the frequency at which the PSD is evaluated; owing to modulation effects of the detector orbits, care must be taken in relating this frequency to the frequency at which a gravitational wave would contribute to $\Delta a_w(t)$; see \secref{signal}.
    The dashed green (respectively, dot-dashed blue) lines show the same PSD quantity, but with all objects that pass within 0.5\,AU of either end of the detector within the mission duration (respectively, the 50 most massive simulated objects in the JPL-SBD) removed from $\Delta a_w(t)$ prior to computing the PSD.	} 
\end{figure}

The results for the one-sided PSD of the windowed baseline-projected differential acceleration, $S^{\textsc{ggn}}[\Delta a_w]$ are shown in \figref{accnResults} for a variety of detector parameters, as listed in \tabref{detectorParams}.
Two sets of detector parameters (simulations No.~1 and No.~2) were chosen to approximate different baselines for detectors located in the  Inner Solar System, and one (simulation No.~3) was chosen to simulate an Outer Solar System mission at approximately the orbit of Neptune (with an optimistic estimate for the baseline).
In all cases, we assumed a mission duration of $T = 10\,$years.
A number of different results are shown in \figref{accnResults} for each detector parameter set, allowing identification of the dominant contributions to the noise at various frequencies: the total noise (solid red), the noise absent the 50 most massive asteroids (dash-dotted blue), and the noise absent all asteroids that pass within 0.5\,AU of either end of the detector baseline during the mission duration (dashed green).

\begin{table}[t]
\begin{ruledtabular}
\caption{\label{tab:detectorParams}%
		Parameters of the single-baseline detector pair utilized in the numerical simulation discussed in \secref{simulation}, and whose results are shown in \figref[s]{accnResults} and \ref{fig:strainResults}.
		The unperturbed detector baseline is $L$, and the detectors are assumed to orbit in the plane of the ecliptic in exactly circular heliocentric orbits at radius $r$; see \eqref{detectorCirc}.
	 }
\begin{tabular}{cll}
Sim.~No.   &   Baseline $L$ [AU] & Orbital radius $r$ [AU] \\ \hline
1                   &   $6.7\times 10^{-3}$ [$=10^{9}$\,m]                     &   1.0                 \\
2                   &   1.0                         &   1.0                 \\
3                   &   30.0                        &   30.0                 \\
\end{tabular}
\end{ruledtabular}
\end{table}

The results for simulations No.~1 and No.~2 are largely insensitive to the assumed start time of the simulation because the detectors complete $\sim 10$ orbits during the simulation; however, the results of simulation No.~3 are sensitive to the exact start time because a detector orbit at $r \sim 30\,$AU takes $T_{\text{orbit}}\sim 165\,$years.
Therefore, our Outer Solar System results are only representative within an $\mathcal{O}(1)$ factor of what a mission with a different assumed start time would experience as the asteroid GGN noise spectrum.

We emphasize also that the frequency at which $S^{\textsc{ggn}}[\Delta a_w]$ is evaluated should not be immediately associated with the frequency of a gravitational waves passing through the Solar System: because we consider the differential acceleration component projected into a rotating baseline direction, a monochromatic GW passing through the Solar System would contribute an effective acceleration to $\Delta a_w(t)$ that has its frequency modulated by the detector orbital frequency, as we discuss in detail in \secref{signal}.
We account for this effect in \secref{signal} in giving limits on how this GGN acceleration noise impacts the detectability of a GW signal; for the moment, we content ourselves with a discussion of $S^{\textsc{ggn}}[\Delta a_w]$ with the frequency content as presented in \figref{accnResults}.

\subsubsection{Inner Solar System} 
The results of the two simulations in the Inner Solar System (No.~1 and No.~2) are broadly similar [up to an amplitude rescaling by the detector baseline; cf.~\eqref{inside}], and can thus be discussed together.

The noise is most severe in the band between $\text{few}\times 10^{-8}$\,Hz and $10^{-7}\,$Hz, drops exponentially quickly by some 10 orders of magnitude between $10^{-7}$\,Hz and $\text{few}\times 10^{-7}$\,Hz, and then hits a much more slowly falling plateau above $\text{few}\times 10^{-7}$\,Hz, falling only another 3 (simulation No.~1) to 6 (simulation No.~2) orders of magnitude by $\text{few}\times 10^{-6}$\,Hz.

As can immediately be seen from the results absent the 50 most massive JPL-SBD objects (masses ranging from $\sim 10^{19}$\,kg to $\sim 10^{21}$\,kg), the dominant noise contribution below $\text{few}\times 10^{-7}$\,Hz are large asteroids that happen to orbit mostly in the plane of the ecliptic (i.e., small inclination) and at heliocentric orbital radii of a few AU.
There is some resolution, given the 10 year simulated mission duration, to resolve effects of individual asteroids in this mass and frequency range (hence the existence of some isolated `bumps' in the PSD).
The ability of remove these 50 asteroids from the acceleration noise (using independently known accurate ephemeris and mass data) would allow some 2--3 orders of magnitude improvement in the noise at these frequencies, as evidenced by the blue dash-dot lines in \figref{accnResults}; however, removal of as large or larger a number of asteroids rapidly becomes challenging, as already discussed in \secref{estimate}.

Nevertheless, this class of large objects in the asteroid belt is unambiguously completely cataloged in the JPL-SBD, and the results in this frequency range are thus highly robust against the addition of the unknown asteroids to the simulation: such new objects simply cannot be in this class, or they would easily already have been discovered.
We also note that because the dominant contribution in this frequency range arises from this specific class of massive, distant, largely in-the-plane, approximately circular-orbit objects, the peak of the noise between $\text{few}\times 10^{-8}$\,Hz and $10^{-7}\,$Hz, and the rapid exponential drop in the noise between $10^{-7}$\,Hz and $\text{few}\times 10^{-7}$\,Hz are both completely expected and in line with the features of the simplified model discussion above in \secref{estimate}, and the more detailed discussion in \appref{multipole}.
Indeed, because the detectors are inside the belt, we would expect [cf.~\eqref{diffAccnExp} or, more specifically, \eqref{specialCaseInside}] the dominant noise to appear around $f \sim 2 f_{\text{det.}} \sim 6\times 10^{-8}\,$Hz, and be suppressed exponentially above this frequency, as higher harmonics are suppressed by powers $\sim (1\,\text{AU} /\text{few\,AU})^j$; this is observed.

On the other hand, at higher frequencies, above $\text{few}\times 10^{-7}$\,Hz, the dramatic decrease in the noise PSD that occurs when the objects that pass closer than 0.5\,AU to either end of the detector baseline at any point during the simulation ($862$ objects in simulation No.~1; $1062$ objects in simulation No.~2) are removed, indicates that the noise above $\text{few}\times 10^{-7}$\,Hz is dominated by these much less massive asteroids (masses ranging from $\sim 10^7$\,kg to $\sim 10^{16}$\,kg, with typical masses around $\sim 10^{9}$\,kg to $\sim 10^{13}$\,kg) that pass much closer to the detectors.
This of course stands to reason, as typical closer-approaching objects will be close flybys%
\footnote{\label{ftnt:notAtruism}%
    As opposed to some special close-approaching objects which could be approximately co-orbiting with the detector.
    }
that exert peak acceleration for a shorter time, leading to higher frequency noise.
However, this class of close-approaching, small objects is not necessarily completely cataloged in the JPL-SBD, and the asteroid GGN result in this frequency range is thus much less robust to the effects of the addition of new, unidentified objects outside the catalog (or, indeed, those that are in the catalog but lack diameter measurements; see \secref{simulationDesc}).
It is thus an open question whether the noise presented in this frequency range is a complete characterization of the full asteroid GGN, or merely a lower bound; we defer detailed investigation of this question to future work (but see \secref{closePasses} for an estimate of the size of the effect). 

\subsubsection{Outer Solar System} 
The results of the Outer Solar System simulation differ from those of the Inner Solar System in important ways.

The noise is now most severe below $10^{-8}$\,Hz, dropping exponentially quickly by some 18 orders of magnitude by $10^{-7}$\,Hz; above $\sim 10^{-7}$\,Hz, the noise is negligibly small.
In other words, the GGN noise here has shifted down in frequency considerably as compared to the Inner Solar System simulations.

As can immediately be seen again from the results absent the 50 most massive JPL-SBD objects, the dominant noise contribution below $\sim 10^{-7}$\,Hz arises from large asteroids that happen to lie in the plane of the ecliptic at heliocentric orbital radii of a few AU.
Removal of these 50 asteroids from the acceleration noise would again allow some 1--2 orders of magnitude improvement in the noise at these frequencies; however, similar comments apply to this procedure as for the Inner Solar System.
Once again, results in this frequency range are robust to the addition of more asteroids to the simulation.

We again note that because the dominant contribution in this frequency range arises from the same specific class of objects of the the Inner Solar System, the peak of the noise below $10^{-8}$\,Hz, and the rapid exponential drop in the noise between $10^{-8}$\,Hz and $10^{-7}$\,Hz are both completely expected and in line with the features of the simplified model discussion above in \secref{estimate}, and the more detailed discussion in \appref{multipole}.
Indeed, because the detectors are now outside the belt, we would expect [cf.~\eqref{diffAccnExp} or, more specifically, \eqref{specialCaseOutside}] the dominant noise to appear around $f \sim f_{\text{ast}} \sim 6\times 10^{-9}\,$Hz [for an assumed 3\,AU asteroid orbit], and to be suppressed exponentially above this frequency, as higher harmonics are suppressed by powers $\sim (\text{few\,AU}/30\,\text{AU})^j$ [actually, with a marginally more severe suppression for the same harmonic as compared to the Inner Solar System case; see discussion in \appref{observations}].
These features are again observed.

No objects in this simulation were identified as passing with 0.5\,AU of either end of the detector in the simulation duration of 10 years, but on the same general grounds as for the Inner Solar System results, we again believe the asteroid GGN at frequencies above $\sim 10^{-7}$\,Hz to be dominated by much less massive asteroids that pass much closer to the detectors.
This noise is negligibly small in this case, but it again remains an open question whether unmodeled close-encounters could increase the noise in this frequency range, as it is highly unlikely that the JPL-SBD captures all small (and thus faint) asteroids at distances of tens of AU from the Sun.\\

While the acceleration asteroid GGN PSDs presented and discussed in this section fully capture the results of our simulation, an additional level of analysis is required to translate these results to limits on the detectable strain amplitude spectral density for a gravitational wave, as a function of the GW frequency. 
We discuss this topic in the next section.

\subsection{A monochromatic gravitational wave signal}
\label{sec:signal}
In order to utilize the asteroid GGN baseline-projected differential acceleration power spectrum we have thus far computed to set limits on the detectable GW amplitude, we must compute how a gravitational wave would give rise to such an acceleration.
We must also take care to apply the same windowing procedure to this signal as to the asteroid GGN.

In the GW frequency range in which we are interested, $f_{\textsc{gw}} \sim 10^{-8}$--$10^{-6}$\,Hz, the GW wavelengths are $\mathcal{O}(10^3$--$10^5\,\text{AU})$, which is much larger than the size of the Solar System, and any orbit or baseline distance of interest to us; cf.~\tabref{detectorParams}.
We can thus analyze the passage of a gravitational wave through the Solar System in the local Lorentz frame of the Sun to a high degree of accuracy (i.e., the $\omega_{\textsc{gw}}L\ll1$ expansion is justified).

To be maximally optimistic in terms of signal detection (i.e., conservative in excluding GW as detectable above the asteroid GGN level), consider a plane gravitational wave incident on the Solar System from the direction exactly perpendicular to the plane of the elliptic, such that the ecliptic plane coincides with a GW phase front; any other direction of incidence will result in a smaller GW-induced test-mass acceleration for a given fixed GW amplitude.
Suppose a test mass is located at $\bm{x} \equiv x \bm{\hat{x}} + y \bm{\hat{y}}$ in the local Lorentz frame of the Sun (chosen such that the elliptic is in the $x$-$y$ plane).
The GW has the effect of inducing the following effective Newtonian acceleration on the test mass \cite{Misner:1974qy}:
\begin{align}
    \begin{pmatrix} 
     \ddot{x} \\
     \ddot{y}
    \end{pmatrix} &=
    \frac{1}{2} 
    \begin{pmatrix}
        \ddot{h}_{+}        & \ddot{h}_{\times} \\[1ex]
        \ddot{h}_{\times}   & - \ddot{h}_{+}
    \end{pmatrix}
    \begin{pmatrix} 
     x \\
     y
    \end{pmatrix},
\end{align}
with no acceleration in the $z$ direction (i.e., direction of the GW motion, perpendicular to the ecliptic), and where $\dot{X}\equiv \partial_t X$, and $h_{+,\times}(t)$ are the gravitational-wave strain waveforms in the $+,\times$ polarizations, respectively.
Assuming that $h_{+,\times}$ are small, we can expand the GW-induced displacement of a test mass as a perturbation around an unperturbed test-mass location as $x = x_0 + \delta x$ (and similarly for $y$), assuming that $|\delta x| / |x_0| \sim \mathcal{O}(h_{+,\times}) \ll 1$ (and similarly for $y$), which gives the following leading equations that govern the perturbations:
\begin{align}
    \begin{pmatrix} 
     \ddot{\delta x} \\
     \ddot{\delta y}
    \end{pmatrix} &=
    \frac{1}{2} 
    \begin{pmatrix}
        \ddot{h}_{+}        & \ddot{h}_{\times} \\[1ex]
        \ddot{h}_{\times}   & - \ddot{h}_{+}
    \end{pmatrix}
    \begin{pmatrix} 
     x_0 \\
     y_0
    \end{pmatrix},
\end{align}
neglecting terms at $\mathcal{O}(h_{+,\times}^2)$.

Considering the action of this GW on the two detectors $A,B$ whose unperturbed positions are given by the circular orbits shown at \eqref{detectorCirc}, and computing the baseline-projected differential acceleration $\Delta a^{\textsc{gw}}$ [as defined at \eqref{DeltaADefn}] induced by the GW, we find that
\begin{align}
    \Delta a^{\textsc{gw}}(t) &\approx -\frac{L}{2} \lb[ \ddot{h}_{+} \cos(2\Omega t) + \ddot{h}_{\times} \sin(2\Omega t) \rb],
    \label{eq:modulation}
\end{align}
where we remind the reader that $\Omega$ is the detector orbital angular frequency (see \secref{estimate} and \appref{diffAccn}).

For monochromatic plane gravitational waves,
\begin{align}
    h_{+,\times}(t,z=0) \equiv h_{+,\times}^{(0)} \cos( \omega_{\textsc{gw}} t + \alpha_{+,\times} ),
    \label{eq:strainWaveform}
\end{align}
where $h_{+,\times}^{(0)}$ are amplitudes, $\alpha_{+,\times}$ are phases, and $\omega_{\textsc{gw}} = 2\pi f_{\textsc{gw}}$, we thus have
\begin{widetext}
\begin{align}
    \Delta a^{\textsc{gw}}(t) &\approx \tfrac{1}{2} L\omega_{\textsc{gw}}^2  h^{(0)}_{+} \cos(\omega_{\textsc{gw}} t+\alpha_{+} ) \cos(2\Omega t) 
   + \tfrac{1}{2} L\omega_{\textsc{gw}}^2 h^{(0)}_{\times} \cos(\omega_{\textsc{gw}}t+\alpha_{\times} ) \sin(2\Omega t)\\
   &= \tfrac{1}{4} L\omega_{\textsc{gw}}^2  h^{(0)}_{+} \lb\{ \cos\Big[ (\omega_{\textsc{gw}} + 2\Omega) t+\alpha_{+} \Big] + \cos\Big[ (\omega_{\textsc{gw}} - 2\Omega) t+\alpha_{+} \Big] \rb\} \nl
   + \tfrac{1}{4} L\omega_{\textsc{gw}}^2 h^{(0)}_{\times}
   \lb\{ \sin\Big[ (\omega_{\textsc{gw}} + 2\Omega) t+\alpha_{\times} \Big] - \sin\Big[ (\omega_{\textsc{gw}} - 2\Omega) t+\alpha_{\times} \Big] \rb\}.
\end{align}
\end{widetext}
We thus see that the GW frequency is modulated by the orbital frequency of the detector pair, as we noted in the previous subsection.

In order to make an apples-to-apples comparison to the asteroid GGN spectrum whose computation we detailed in the previous two subsections, we apply the same window function $w(t)$ to this signal as to the noise:
\begin{align}
    \Delta a_w^{\textsc{gw}}(t) & \equiv w(t) \cdot \Delta a^{\textsc{gw}}(t),
\end{align}
where $w(t)$ is defined at \eqref{windowFunc}.
We can then sample $\Delta a^{\textsc{gw}}_w(t)$ at the same times $t_n = n\Delta t = nT/\mathcal{N}$ ($n=0,\hdots,\mathcal{N}-1$) as for the asteroid GGN, and compute the (windowed) DFT, $\widetilde{\Delta a_w^{\textsc{gw}}}(f_k)$.

It is then straightforward to show that the signal-to-noise ratio (SNR) $\rho$ for the detection of this GW signal above the asteroid GGN noise floor in a standard matched-filter search is given by \cite{Maggiore:2007zz,Moore:2014lga}
\begin{align}
    \rho^2 &= \frac{1}{T} \frac{ | \widetilde{\Delta a_w^{\textsc{gw}}}(f_0) |^2}{S_0^{\textsc{ggn}}[\Delta a_w]} +  \frac{1}{T} \frac{| \widetilde{\Delta a_w^{\textsc{gw}}}(f_{\mathcal{N}/2}) |^2}{S_{\mathcal{N}/2}^{\textsc{ggn}}[\Delta a_w]} \delta_{0,\mathcal{N}\!\!\!\!\!\mod 2} \nl
    + \frac{4}{T} \sum_{k=1}^{\lb\lfloor (\mathcal{N}-1)/2 \rb\rfloor} \frac{ | \widetilde{\Delta a_w^{\textsc{gw}}}(f_{k}) |^2 }{ S_k^{\textsc{ggn}}[\Delta a_w] },
    \label{eq:matchedFilter}
\end{align}
where the term at $k=\mathcal{N}/2$ exists only for even $\mathcal{N}$, as indicated by the Kronecker delta, and where we have adapted the continuous results of \citeR[s]{Maggiore:2007zz,Moore:2014lga} to the discrete-sampling case at hand.

Treating each of the $+$ and $\times$ polarization cases in turn (assuming in each case that the other polarization is absent), at each GW frequency we use \eqref{matchedFilter} to find the values of $h_{+,\times}^{(0)}$ such that $\rho = 1$; we denote these values of the detectable strain amplitude by $\hat{h}_{+,\times}^{(0)}(f_{\textsc{gw}})$. 
We convert each of these results to an effective GGN strain amplitude spectral density $\sqrt{S_{+,\times}^{\textsc{ggn}}[h]}$; i.e., the strain amplitude spectral density that, given the signal size already found to yield an SNR of 1 in the matched-filter search as applied to $\Delta a_w$ (note: windowed), would result in an SNR of 1 in a matched-filter search on the GW strain, $h$ (note: unwindowed). 
Since the monochromatic signal is narrow-band, the appropriate conversion [assuming a non-edge case; cf.~\eqref{matchedFilter}] is
\begin{align}
    1 \sim \frac{4}{T} \frac{\Big|\tilde{h}_{+,\times}\big[f_{\textsc{gw}};\hat{h}_{+,\times}^{(0)}(f_{\textsc{gw}})\big]\Big|^2}{S_{+,\times}^{\textsc{ggn}}[h](f_{\textsc{gw}})} \sim \frac{T\cdot |\hat{h}_{+,\times}^{(0)}(f_{\textsc{gw}})|^2}{S_{+,\times}^{\textsc{ggn}}[h](f_{\textsc{gw}})}\label{eq:ShNoWindow} \\
    \Ra \sqrt{S_{+,\times}^{\textsc{ggn}}[h](f_{\textsc{gw}})} \equiv \sqrt{T} \cdot \hat{h}_{+,\times}^{(0)}(f_{\textsc{gw}}),\label{eq:ASDconv}
\end{align}
where $\tilde{h}_{+,\times}[f_{\textsc{gw}};\hat{h}_{+,\times}^{(0)}(f_{\textsc{gw}})] \sim (T/2) \hat{h}_{+,\times}^{(0)}(f_{\textsc{gw}})$ is the unwindowed, single-signal-bin DFT result for the signal \eqref{strainWaveform} with $h^{(0)}_{+,\times} = \hat{h}_{+,\times}^{(0)}(f_{\textsc{gw}})$; cf.~\eqref{singleCosine}.%
\footnote{\label{ftnt:approx}%
	This is of course only strictly speaking correct if the frequency of the GW is at an exact DFT frequency.
	To the extent that there are any approximations interposed in this motivating derivation, we consider \eqref{ASDconv} as the exact definition of $S_{+,\times}^{\textsc{ggn}}[h](f_{\textsc{gw}})$, with $\hat{h}_{+,\times}^{(0)}(f_{\textsc{gw}})$ defined so as to make $\rho=1$ in \eqref{matchedFilter}.
	This removes any ambiguity or approximation in this discussion.
	}

We note that this somewhat convoluted procedure is designed to appropriately fold in both the effects of the GW signal modulation that is shown at \eqref{modulation}, and the effects of the windowing.
Absent the modulation effects, we would obtain the (mostly) expected result $\sqrt{S_{+,\times}^{\textsc{ggn}}[h](f_{\textsc{gw}})} \sim 2.3 \sqrt{S^{\textsc{ggn}}[\Delta a_w](f_{\textsc{gw}}) / (\omega_{\textsc{gw}}^2 L/2)^2}$.
The numerical constant $\sim 2.3$ arises because of windowing: there is a mismatch between $S^{\textsc{ggn}}[\Delta a_w]$ being defined in terms of the windowed $\Delta a_w$, and $S_{+,\times}^{\textsc{ggn}}[h]$ being defined without regard to a window function on the signal at \eqref{ShNoWindow}; this was done so that the latter result is more easily interpretable as the usual effective strain noise to employ in deciding whether a signal is detectable without having to be concerned with any technicalities of our analysis that arise from windowing.
This numerical constant of $\sim 2.3$ would instead be exactly 1 if no windowing was necessary anywhere.%
\footnote{\label{ftnt:characteristic}%
    We also note that the conversion at \eqref{ASDconv} yields the usual rule for the detectable \emph{broadband} GW characteristic strain that is obtained for a monochromatic signal with detectable strain amplitude $\hat{h}_{+,\times}^{(0)}$: $h_c^2 \sim {f_{\textsc{gw}} \cdot S_{+,\times}^{\textsc{ggn}}[h](f_{\textsc{gw}})} \sim {N_{\text{cycles}}}\cdot [\hat{h}^{(0)}_{+,\times}]^2$, where $N_{\text{cycles}} = f_{\textsc{gw}} T$ is the number of GW cycles observed during a time $T$.} %

We show the results for the effective GGN strain amplitude spectral density, $S_{+,\times}^{\textsc{ggn}}[h]$, in \figref{strainResults}, for the same three sets of detector parameters outlined in \tabref{detectorParams}.

\subsubsection{Discussion of strain results}

\begin{figure*}[p]
\includegraphics[width=0.67\textwidth]{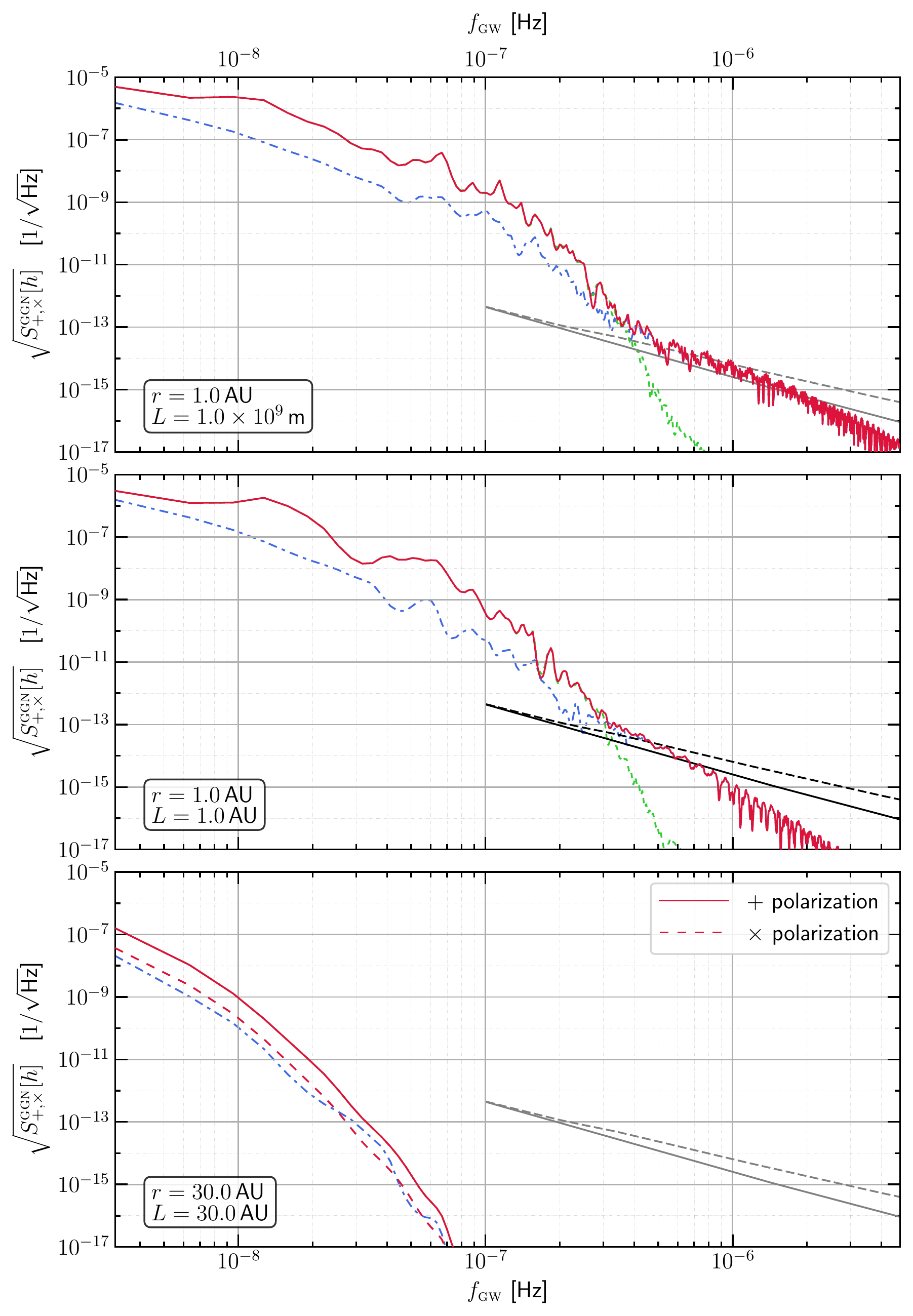}
\caption{\label{fig:strainResults}%
    Gravitational wave strain amplitude spectral density sensitivity corresponding to a signal-to-noise ratio of one in \eqref{matchedFilter} for a gravitational wave of frequency $f_{\textsc{gw}}$, incident on the Solar System from perpendicular to the plane of the ecliptic, for a pair of detectors at heliocentric radius $r$ with a baseline $L$, and subject to the GGN acceleration induced by objects in the JPL-SBD that is shown in \figref{accnResults}.
    The solid red%
    \footnote{ The results for $r=L=30.0\,$AU are presented separately for the $+$ (solid red) and $\times$ (dashed red) polarizations; see text for discussion. } %
    lines show the strain ASD $\sqrt{ S^{\textsc{ggn}}[h]}$ as defined at \eqref{ASDconv} and computed per the procedures outlined in \secref{signal}.
    These results account only for those objects in the JPL-SBD with an explicitly supplied diameter (see \secref{simulationDesc} for discussion), and for the Inner Solar System baselines they may thus represent an underestimate of the noise at higher frequencies ($f \gtrsim \text{few}\times 10^{-7}\,$Hz); see the detailed discussion around \figref{higherClosePassNoise} in \secref{closePasses}.
    We assume a mission duration of $T = 10\,\text{years}$ and $\mathcal{N}=3000$ sampling points in the numerical simulation, and results are displayed for frequencies $f \in [ \Delta f , f_{\text{Nyquist}}]$.
    The dashed green (respectively, dot-dashed blue) lines show the same strain ASD quantity, but for the cases where the asteroid GGN noise contributions from all objects that pass within 0.5\,AU of either end of the detector within the mission duration (respectively, the 50 most massive simulated objects in the JPL-SBD) are all removed \emph{from the GGN noise} prior to computing the SNR.
    Also shown in the middle panel by the solid (respectively, dashed) black lines are the $\mu$Ares sensitivity curves given the expected instrumental (respectively, instrumental plus astrophysical) noise for that proposed mission concept \cite{Sesana:2019vho}.
    Note that the $\mu$Ares curves assume a three-detector configuration with baselines of $\sim 3\,$AU \cite{Sesana:2019vho}, and so are not directly comparable to any of our scenarios absent some $\mathcal{O}(1)$-factor adjustment.
    Nevertheless, we show these curves exactly as extracted from \citeR{Sesana:2019vho} for the closest approximant that we simulated: $r=L=1.0\,$AU (note that we also show the same curves as solid or dashed grey lines on the other two panels; this however is  purely for comparative visual reference).
    } 
\end{figure*}

As for the acceleration results discussed previously, the Inner Solar System simulations (No.~1 and No.~2) yield broadly similar results at low frequencies, although the longer baseline result yields improved detection performance at higher frequencies by up to 2 orders of magnitude.
We again note however that the high-frequency ($f\gtrsim \text{few} \times 10^{-7}\,$Hz) portion of these results may constitute only a lower bound on the noise floor, and is sensitive to undetected, close-approaching objects not captured in the JPL-SBD.

Nevertheless, taking these results at face-value, comparison%
\footnote{\label{ftnt:muAreaComparison}%
    	We note that the $\mu$Ares noise curves are obtained in \citeR{Sesana:2019vho} using an assumed three-detector configuration of detectors with $\sim 3\,$AU baselines.
    	The comparisons of our results to the $\mu$Ares curves in \figref{strainResults} should not be understood as being numerically precise.
    	Our $r=L=1.0\,$AU baseline results are the most comparable, but the comparison should be understood to be valid only up to an $\mathcal{O}(1)$ factor.
    } %
of the Inner Solar System results to the limiting strain ASD noise curves for the proposed $\mu$Ares detector array \cite{Sesana:2019vho}, which are shown as solid (respectively, dashed) black lines in \figref{strainResults} taking into account only expected instrumental (respectively, both instrumental and astrophysical) noise for that proposed mission concept, indicates that asteroid GGN becomes a noise source comparable to instrumental and astrophysical noise for frequencies $f_{\textsc{gw}} \sim \text{(few)} \times 10^{-7}\,$Hz and---by orders of magnitude---the dominant (and therefore limiting) noise source at lower frequencies.%
\footnote{\label{ftnt:conversionmuAres}%
	The $\mu$Ares noise curves in \citeR{Sesana:2019vho} are given in terms of the characteristic strain $h_c$; we have converted to strain ASD using $\sqrt{S[h]} = h_c / \sqrt{f}$.
	} %
It is important to note that even the removal of $\mathcal{O}(50)$ of the most massive objects in the JPL-SBD would not substantially alter these qualitative low-frequency conclusions.
However, at higher frequencies, the amplitude of the asteroid GGN presented here appears to be (marginally, in the short-baseline case of simulation No.~1) small enough to not cut into interesting parameter space, subject to the caveat noted above regarding undetected, close-approaching objects.

On the other hand, the results for the Outer Solar System simulation (No.~3) show that the asteroid GGN floor is a much less severe problem for such a mission. 
While the $\mu$Ares noise curves are only given for $f_{\textsc{gw}} \gtrsim 10^{-7}$\,Hz in \citeR{Sesana:2019vho}, na\"ively extrapolating those noise curves to lower frequencies indicates that asteroid GGN would only become a problematic noise source for detection of GW with $f_{\textsc{gw}} \lesssim 1$--$2\times 10^{-8}\,$Hz.

As we noted in \secref{results}, because the detectors complete only a fraction $\sim 1/16$ of a full $r= 30\,$AU orbit in the simulated 10 year mission duration, the Outer Solar System results we present here are somewhat sensitive to the exact start time of the simulation, and also to the exact assumed polarization of the GW at the start time of the simulation and its relation to the orbital phase offset of the detectors at that time. 
For instance, taking into account the qualification about the start time of the simulation in footnote \ref{ftnt:offset}, the result at \eqref{modulation}, and the effect of our chosen window function [\eqref{windowFunc}] to emphasize the relative importance of the signal in the middle of the simulation duration as opposed to near the start and end, it is the case that the $+$ polarization has, for the same amplitude GW signal, a smaller detector response by roughly a factor of 10 as compared to the $\times$ polarization over the particular duration of the 10 year assumed mission lifetime simulated (effectively, for the $+$ case, the window function happens to localize a region of the signal response which is near a node in the modulation of the GW by the orbital motion of the detectors).
Of course, modifying the simulation start time or assumed mission duration would modify the relative limits on the two different polarizations; the difference between the $+$ and $\times$ results in the lower panel of \figref{strainResults} is representative of the magnitude of the changes one might expect by varying such parameters.
Such dependence is of course absent for the Inner Solar System results, as the detectors in that case complete ten full orbits during the simulation, which effectively reduces the dependence of the results on the temporal location of nodes in the detector response, etc.

For completeness, we note that we have assumed in all cases (both Inner and Outer Solar System) that $\alpha_{+,\times}=0$ at the start of the simulation.

\subsection{Discussion of unmodeled close passes}
\label{sec:closePasses}
We have already noted that the JPL-SBD is likely incomplete for small asteroids that pass close to the detector network, and that such close passes are likely to contribute additional high-frequency noise above \linebreak$\sim\text{(few)} \times 10^{-7} \, \text{Hz}$.
In this subsection, we give a preliminary estimate for the size of that noise.

Consider that a close pass of an asteroid to the detector network can be approximated for the relevant portion of the motion during which it exerts peak acceleration on one or both of the detectors as moving in a straight line, with $v\sim 30$\,km/s being a typical relative speed between the asteroid and the detector at either end of the baseline, provided we consider an Inner Solar System detector network.
As shown in, e.g., \citeR{LISA-Pre-Phase-A}, such motion contributes to accelerations dominantly at frequencies such that $\omega b/v \sim 1$, or $f\sim v/(2\pi b)$ where $b$ is the impact parameter for the fly-by.
As we will see, we are most concerned about this close-encounter noise for frequencies $f\sim \mu$Hz, which yields $b \sim v/(2\pi f) \sim 5\times 10^9\,\text{m} \sim 3.1\times 10^{-2}\,$AU.
Throughout this subsection, we take $f \sim \mu$Hz as an example to estimate how much our calculation could be underestimating the effect of close passes.

\begin{figure}[t]
\includegraphics[width=\columnwidth]{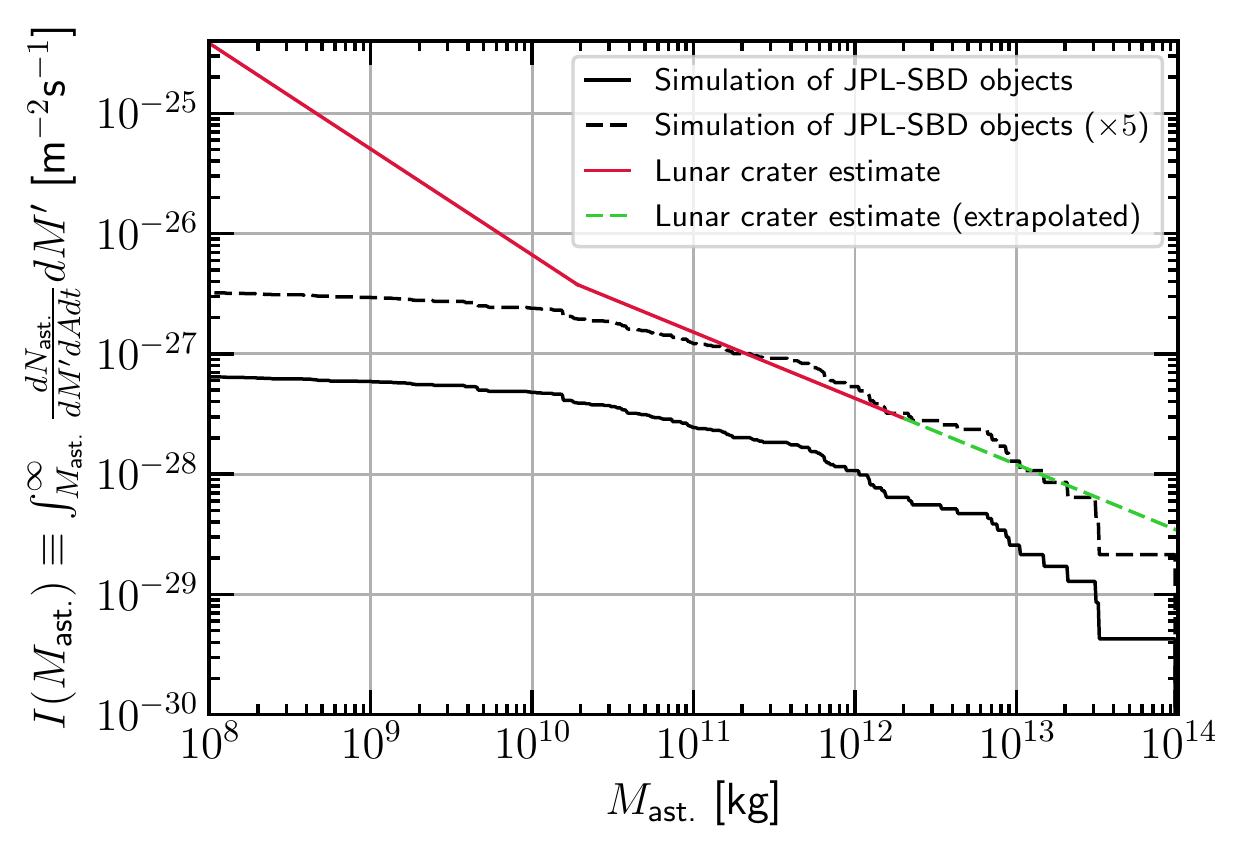}
\caption{\label{fig:closepassflux}%
    The integrated flux of asteroids with masses greater than or equal to mass $M_{\text{ast}}$ passing near the Earth, as a function of the asteroid mass.  
    The solid black line shows the results from our simulation of the JPL-SBD objects. For reference we have also plotted 5 times that line (black dashed). 
    The solid red line shows the estimate from Lunar impact craters \cite{Shoemaker:1983qae, LISA-Pre-Phase-A}, while the dashed green curve is an extrapolation of those data to higher mass.
	} 
\end{figure}

First we estimate how many asteroids may be missing from the JPL-SBD (or, at least, those objects in the JPL-SBD for which a diameter $d_i$ is available to allow a mass determination to be made), and hence from our simulation; see also the discussion in \secref{omittedObjects}.
Data from Lunar impact craters \cite{Shoemaker:1983qae} can be used \cite{LISA-Pre-Phase-A} to estimate an upper bound on the flux of asteroids in the vicinity of a $\sim 1\,$AU orbit in the Inner Solar System, averaged over the age of the Lunar surface.
\figref{closepassflux}~provides a comparison of this flux to the flux in our simulation of the JPL-SBD objects.
Note that for large asteroids with mass above $\sim 10^{11} \, \text{kg}$, the Lunar impact estimate appears to match closely to 5 times that from those JPL-SBD objects we simulate.
Below this mass, the gap between the Lunar impact estimate and our simulation of the JPL-SBD objects grows larger.
For example at an asteroid mass of $3\times 10^{10} \, \text{kg}$, the ratio of fluxes is about a factor of $\sim 10$.

It is generally expected that our current observations could miss significant numbers of asteroids with diameters below about 1\,km (mass very roughly around $10^{12} \, \text{kg}$), but should find most of the asteroids above this size.
For example, both the diameter ($d_i$) distribution for the objects in the JPL-SBD that we do model, and the estimated diameter ($\bar{d}_{H,i}$) distribution for the objects in the JPL-SBD that we do not model, exhibit a decrease in the number of objects with (estimated) diameters smaller than 1\,km, as compared to those with (estimated) diameters around 1\,km.
The most likely explanation then for the above observations about the low-mass discrepancy in flux between our simulation and the Lunar impact estimate in \figref{closepassflux} is that the JPL-SBD as a whole is simply missing significant numbers of asteroids with masses below about $\sim 10^{11} \, \text{kg}$.

Two possible reasons suggest themselves for the factor-of-5 discrepancy that exists at higher asteroid masses between the flux estimate based on the objects in our simulation of the JPL-SBD objects, and the Lunar impact estimate:
(1) this could be reflective of the roughly 5-to-1 ratio of unsimulated-to-simulated JPL-SBD objects in this work; indeed, as our discussion in \secref{omittedObjects} makes clear, the unsimulated population is generally a population of closer-passing asteroids that are typically in the $\bar{d}_{H,i} \sim 1$--$10\,$km class. 
Their omission from the simulated near-Earth asteroid flux could thus easily account for the discrepancy; however, a detailed quantitative accounting for this extra flux would require detailed orbital tracking for the unmodeled class, would be most appropriate in the follow-up simulation work discussed below; and/or
(2) the JPL-SBD is based on actual observations of asteroids at the present epoch, whereas the Lunar data are averaged over the age of the Lunar surface.
But the true number of asteroids in the Solar System may be changing in time (as could the mass distribution, owing to asteroid dynamics), so that the flux at the present epoch could be smaller than the average flux measured over the age of the Lunar surface.
To address case (1), we make the conservative guess (i.e., worst-case scenario), that we are actually missing from our results the entire difference between the Lunar impact estimate for the flux and the flux estimated from our simulation of JPL-SBD objects (i.e., the entire gap between the \emph{solid} black line and the red/green line in \figref{closepassflux}).
To address case (2), we make the optimistic guess that at the present epoch we are simply missing a flux equal to the \emph{residual} difference between the Lunar impact estimate and \emph{5 times} the JPL-SBD object flux in our simulation (i.e., the gap between the black \emph{dashed} line and red/green line in Fig.~\ref{fig:closepassflux}).
We give estimates for both of these scenarios in this section.

\begin{figure}[t]
\includegraphics[width=\columnwidth]{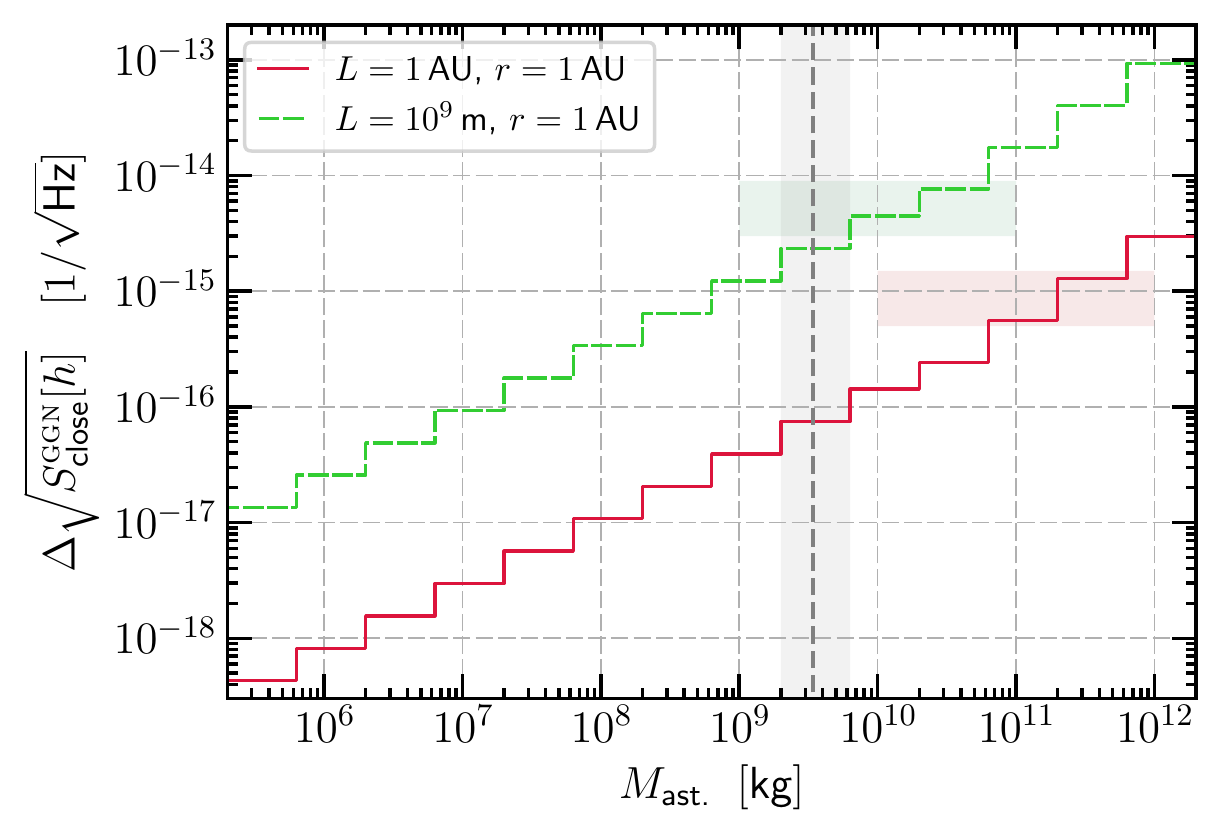}
\caption{\label{fig:closepassstrain}%
    A rough analytic estimate of the contribution of close passes to the GW strain ASD noise at $f \sim \mu$Hz (red solid and green dashed lines, for the two Inner Solar System baselines as indicated in the legend).
    The quantity $\Delta \sqrt{S^{\textsc{ggn}}_{\text{close}}[h]}$ is the contribution to the strain ASD noise from each half-decade-wide mass bin indicated, per the estimate outlined in the text.
    Also shown are lightly shaded, horizontal, red and green colored bands that give the approximate noise levels in our simulation of the JPL-SBD objects (cf.~\figref{strainResults}), for the respective baselines; the vertical width of these bands gives an indication of the amount by which the noise level in the simulations varies in the vicinity of $f_{\textsc{gw}}\sim\mu$Hz (note that the horizontal positioning of these bands on this plot is for visualization purposes only and is not intended to convey any information about a mass range; these colored bands convey information only about the strain ASD).
    The vertical gray shaded band indicates the `cutoff' mass bin for these estimates (see discussion in the text), while the vertical dashed gray line is the average asteroid mass $M_{\text{avg}}$ in this bin.
	} 
\end{figure}

Next we estimate how much these missing asteroids could increase the noise from close passes, which are the dominant contribution to the higher frequency end of the curves in Fig.~\ref{fig:strainResults}. 
Since we have an estimate of the factor by which we may be under-counting asteroids, we attempt to estimate parametrically how the close-pass noise will scale with this number.

First we estimate how much noise each piece of the asteroid mass distribution contributes by itself (we consider slices of the asteroid mass distribution half an order of magnitude in width); see Fig.~\ref{fig:closepassstrain}.  
For this estimate, we use the asteroid number distribution implied by the Lunar impact data.
Because the asteroids contribute stochastically to the acceleration noise, we might expect that the effective acceleration noise contribution exerted on either proof mass in the detector by asteroids in any slice can be estimated by multiplying the acceleration noise from the average asteroid (mass $M_{\text{avg}}$) in that slice by the square root of the number of asteroids in that slice, $\sqrt{ N_\text{ast}}$.%
\footnote{\label{ftnt:averageMass}%
    That is, we can consider each slice in the mass distribution to exert the same acceleration as a single object with an effective mass $M_{\text{eff.}} \equiv M_{\text{avg}} \sqrt{ N_\text{ast}} \equiv M_{\text{slice}} / \sqrt{N_{\text{ast}}}$, where $M_{\text{slice}}$ is the total asteroid mass in the relevant slice.
    }
The acceleration from an asteroid of mass $M_{\text{avg}}$ at a typical distance $\sim b$ would contribute an acceleration on either detector in the baseline of order $a \sim G_N M_{\text{avg}} / b^2$. 
We will assume a worst-case scenario by doubling this estimate to account for acceleration on the proof masses at both ends of the baseline (note that this intentionally omits any accounting for orientation effects); however, if $L<b$, we take a (tidal) baseline suppression by a factor of $L/b$.
That is, we estimate the baseline-projected differential acceleration noise from each slice to be $\Delta a \sim \big(2 G_N M_\text{avg} \sqrt{ N_\text{ast}}  / b^2 \big) \times \text{min}[ 1, L/b]$.

Roughly, and conservatively, we can assume that this will give rise to a strain contribution of 
\begin{align}
h_c &\sim \frac{ \Delta L}{ L} \sim \frac{1}{2} \frac{ (\Delta a) T^2 }{ L } \\
&\sim \frac{ 4\pi^2 G_N M_\text{avg} \sqrt{ N_\text{ast}} }{ L v^2}  \times \text{min}\lb[ 1,  \frac{2\pi L}{vT} \rb] \\
&\sim \frac{ 4\pi^2 G_N M_{\text{slice}} }{  v^2 L \sqrt{N_{\text{ast}}} } \times \text{min}\lb[ 1,  \frac{ 2\pi L}{vT} \rb],
\end{align}
 where we took $T = 1/f = 2\pi b/v \sim 10^6\,$s as the relevant timescale.
From this $h_c$ estimate, we extract the strain ASD contribution as $\Delta \sqrt{S^{\textsc{ggn}}_{\text{close}}[h]} \sim h_c / \sqrt{f}$.
The results of this calculation are shown in Fig.~\ref{fig:closepassstrain} as a function of the asteroid mass, for both of our Inner Solar System baselines.
The vertical shaded gray band shows the asteroid mass slice for which roughly one asteroid passes within a distance $b$ in time $T\sim 1/f$ (we call this the `cutoff' mass); the vertical dashed line gives the average asteroid mass for this slice.

Fig.~\ref{fig:closepassstrain} demonstrates that the close pass noise is dominated by the largest asteroids.
The noise curves in Fig.~\ref{fig:closepassstrain} do not apply for masses much greater than the vertical dashed line that marks the average mass of the asteroids in the slice for which on average approximately one asteroid passes within a distance $b$ in a time $T$, because such asteroids pass near the detector only rarely and thus are not the same kind of continuous noise as would be confounding for a GW at frequency $f_{\textsc{gw}}\sim 1/T$ (i.e., the curves in \figref{closepassstrain} do not apply above the `cutoff' mentioned above).
So, as an analytic estimate for the noise, we take the level of the strain ASD noise curve in \figref{closepassstrain}, evaluated at the cutoff mass.
Of course in reality the answer should be some integral up to (and possibly somewhat above) the cutoff mass.
Nevertheless, comparing to the shaded horizontal colored bands in \figref{closepassstrain} we see that our analytic estimate in the dominant mass slice is within about an order of magnitude of our simulation results, but that the former underestimates the latter.
The underestimate would be partially compensated by taking an integral; we will not be concerned with this because we are actually only looking for the parametric dependence of this noise level on the asteroid flux.  

We thus see from \figref{closepassstrain} that the noise is dominated by the largest asteroids below the cutoff, with masses around $\sim 3\times 10^{9} \, \text{kg}$.
This is important, because it means we only have to estimate by how much asteroids in this mass class are being under-counted in our numerical simulation as compared to the Lunar impact flux estimate, and the impact of increasing the flux of asteroids in this mass class by the factor required to match the Lunar impact flux estimate, in order to understand the possible range of factors by which we could be underestimating the strain ASD noise.

\begin{figure}[t]
\includegraphics[width=\columnwidth]{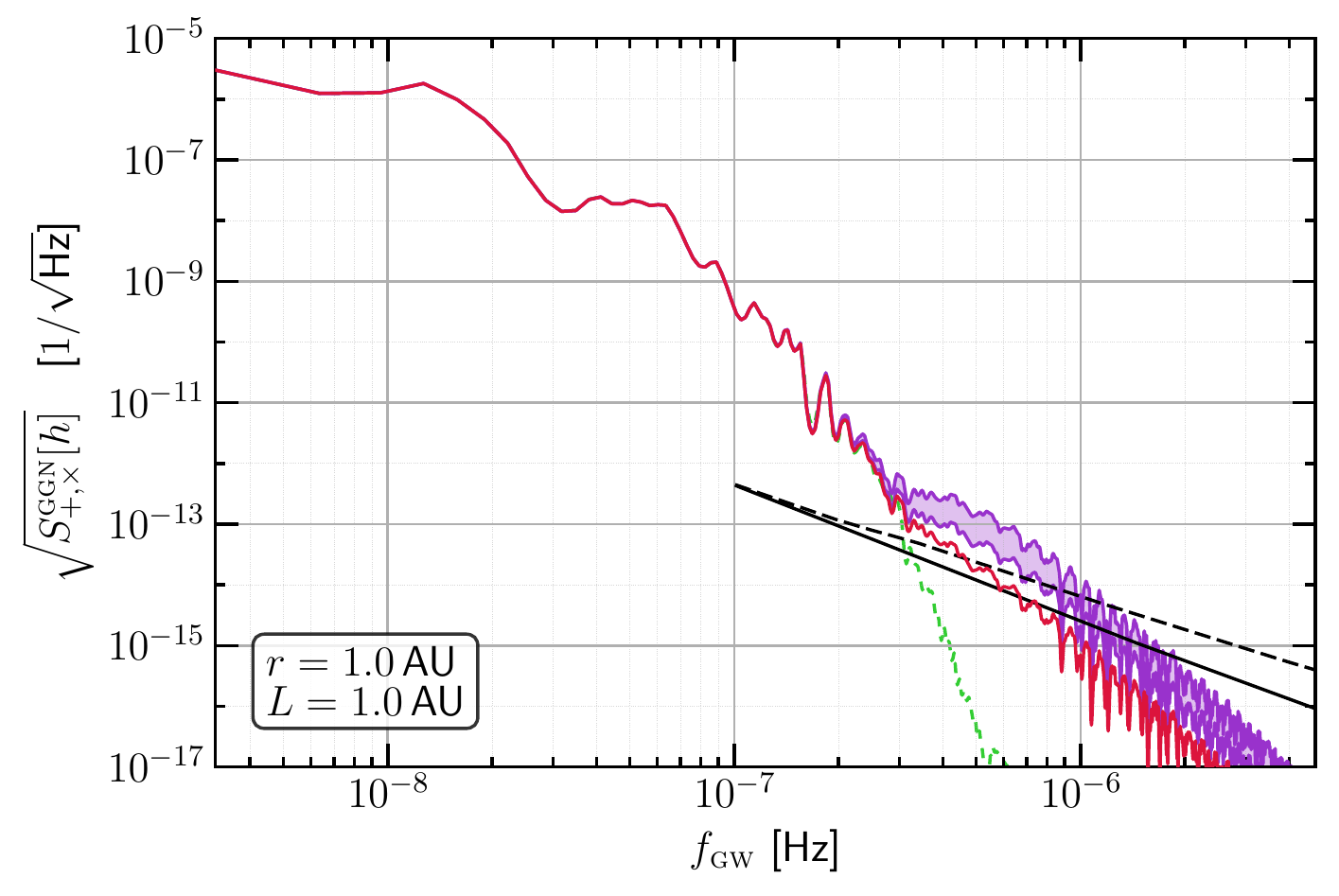}
\caption{\label{fig:higherClosePassNoise}%
    Schematic estimate of the possible additional high-frequency noise from close-passing objects not contained in our simulation.
    The solid red and green dashed lines are our simulation results from the middle panel of \figref{strainResults}, with or without asteroids that pass with 0.5\,AU of either end of the detector included, respectively.
    The black solid and dashed curves are the $\mu$Ares curves described in the caption of \figref{strainResults} \cite{Sesana:2019vho}.
    To indicate the additional high-frequency close-pass strain ASD noise estimate discussed in the text, we plot the strain noise ASD from simulation increased by a factor of 3--10 for $f_{\textsc{gw}} \gtrsim 3\times 10^{-7}\,$Hz  (lower and upper edges of the shaded purple band, respectively). 
    We stress that this purple band is a rough estimate for this additional noise, and is not a final answer; this noise contribution warrants further study in future work.
    Note also that, in this plot, we have assumed a simple (functionally smoothed) turn-on of this additional noise around $f_{\textsc{gw}} \sim 3\times 10^{-7}\,$Hz; the exact shape of the curve around this frequency would also need to be more carefully computed.
	} 
\end{figure}

To make an estimate of how the close-pass noise would increase if we increased%
\footnote{\label{ftnt:scalingUp}%
    The reader may question why we are \emph{increasing} by a factor of $x$ the result in \figref{closepassstrain} that is based on the \emph{Lunar impact} flux data, which already exceeds the flux of JPL-SBD objects in our simulation.
    Here we are only asking for the parametric scaling of this strain estimate with changing flux normalization under the assumption that the number of asteroids in each slice scales with mass as in the Lunar impact data flux estimate.
    For the purposes of obtaining this parametric scaling, the starting point from which the overall flux is rescaled upward is not relevant (at least in the asteroid mass range relevant to us for these estimates).
    We use this parametric scaling to extract a `correction factor' under the assumption of a flux increase by a factor of $x$ that we then apply to increase our \emph{simulation} strain ASD noise results; this procedure is consistent with the relative flux normalizations in \figref{closepassstrain}.
    } %
the asteroid number flux by a factor $x$, we perform the same calculation as in \figref{closepassstrain} but with an increased asteroid number flux.
This moves the strain ASD curves up vertically by $\sqrt{S[h]} \propto x^{0.5}$, but also increases the cutoff mass (i.e., the heaviest asteroid passing within a distance $b$ within time $T$). 
Taking all this into account, we find for example that if the overall number of asteroids is increased by a factor of $10$ then the cutoff mass increases by roughly a factor of $\sim 10$ (i.e., the vertical dashed gray band in \figref{closepassstrain} moves to the right by a factor of 10, or 2 bins on that plot), and we find that the overall high-frequency close-pass strain ASD noise level should thus be taken to increase%
\footnote{\label{ftnt:scalingWithMass}%
        That this should be the scaling is clear from our estimate above that $\sqrt{S^{\textsc{ggn}}[h]} \propto \Delta a \propto M_{\text{avg}}\sqrt{N_{\text{ast}}}$.
        Since we are fixing $M_{\text{avg}}$ to be the average asteroid mass in the bin for which $N_{\text{ast}} \sim 1$, $\sqrt{S^{\textsc{ggn}}[h]} \propto M_{\text{avg}}$, so if the average mass in the cutoff bin goes up by a factor of 10, so too does the strain ASD noise.
    } %
by a factor of $\sim 10$.
Recalling the earlier discussion that in the worst case scenario we called (1), the number of relevant-mass asteroids may actually be larger than the JPL-SBD objects in our simulation by a factor of at most $\sim 10$ (cf.~\figref{closepassflux} in the vicinity of $M_{\text{ast}}\sim 3\times 10^{10}\,$kg), our worst-case estimate is that the high-frequency close-pass noise may be larger than our simulation answer by a factor of at most $\sim 10$ around $f_{\textsc{gw}}\sim\mu$Hz.
However, in the optimistic scenario we called (2), the number of asteroids around $M_{\text{ast}} \sim 3 \times 10^{10}\,$kg increases by only a factor of $\sim 2$ when comparing our simulation flux to the Lunar impact flux estimate because a factor of $\sim 5$ in the difference would be due to a mismatch between present-day and historical fluxes of asteroids near the Earth (see the discussion around \figref{closepassflux} above); in this case, we find that the strain ASD noise would only increase by a factor of $\sim 3$ around $f_{\textsc{gw}}\sim\mu$Hz.
Assuming these multiplicative factor increases to be uniform across the decade of frequencies $f_{\textsc{gw}} \sim \text{(few)} \times 10^{-7}\,$Hz--$\text{(few)} \times 10^{-6}\,$Hz, we sketch this level of additional strain ASD noise (i.e., a factor of 3--10 larger than our simulation result) as the shaded purple band in \figref{higherClosePassNoise}.

While we have attempted to conservatively estimate the range in which the high-frequency close-pass noise may lie, we note that there is significant uncertainty in our estimate, as indicated by the shaded band in \figref{higherClosePassNoise}.
More generally, our estimates in this subsection for the how much the true high-frequency close-pass strain ASD noise exceeds the level in our simulation results should be considered to be an informed, but rough, order-of-magnitude guess.
The salient point to draw from this discussion is that an accurate calculation of this high-frequency noise from close passes of less massive asteroids may be relevant for experimental concepts aiming at the $\mu$Hz frequency range, such as $\mu$Ares \cite{Sesana:2019vho}.
One approach to improve understanding on this point would be to perform a simulation of the class of smaller, closer-passing asteroids that we left unmodeled in this work, by making mass estimates for those asteroids based on the diameter estimates $\bar{d}_{H,i}$ obtained from individual asteroid absolute magnitudes and the average asteroid geometric albedo, as discussed in \secref{simulationDesc}.
We defer such calculation to future work.

\section{Conclusions}
\label{sec:conclusions}
In this paper, we calculated the acceleration noise on a gravitational wave detector employing local proof masses that arises from the gravitational influence of asteroids in the Solar System.
Our main results are shown in \figref[s]{strainResults} and \ref{fig:higherClosePassNoise}.
\figref{strainResults} shows the conservative minimum noise floor arising only from asteroids whose properties are well-measured and reported in the JPL-SBD catalog.
\figref{higherClosePassNoise} attempts to estimate the full noise including all asteroids in the Solar System, and in particular how the high-frequency tail of the results depends on unmodeled, smaller asteroids passing closer to the detector.

Our analysis shows that gravity gradient noise from asteroids in the asteroid belt will severely limit the sensitivity of any gravitational wave detector that uses local proof masses in the frequency band below $\sim\text{(few)}\times10^{-7}$\,Hz as long as the detector baseline is located in the Inner Solar System; see \figref{strainResults}.

At higher frequencies, the noise is dominated by close encounters between asteroids and the detector.
With the objects in the JPL-SBD that we have simulated, we find this effect to be sufficiently small to permit gravitational wave detection above $\sim \mu$Hz; see \figref{strainResults}.
However, this analysis is incomplete as the JPL-SBD objects we have utilized in our simulations likely do not represent a complete sampling of objects that are smaller than $\sim$ km.
Using a near-Earth asteroid flux estimate obtained from Lunar impact data \cite{Shoemaker:1983qae,LISA-Pre-Phase-A} we found reasonable estimates for the contribution of these objects to the asteroid GGN yielded an increase in the high-frequency noise around $\sim\mu$Hz by a factor of $\sim 3$--$10$ as compared to the high-frequency noise we have explicitly computed from JPL-SBD objects; see \figref{higherClosePassNoise}.
This noise contribution thus warrants further detailed study, especially to establish the viability of proposed detectors in this frequency band.
We leave this for future work. 

We also note that while asteroid gravitational influences on a GW detector constitute noise from the perspective of a GW detection, these effects can also be viewed as a positive signal from the point of view of studies of asteroids.
There are potentially interesting questions regarding the dynamics of asteroids, and their mass and spatial distributions that could be addressed by such low-frequency measurements.

The results of this paper in the frequency band below $\sim\text{(few)}\times10^{-7}$\,Hz raise an interesting question: given that one cannot straightforwardly use local test masses in the Inner Solar System, how could gravitational waves in this frequency band be detected? 
We have shown that placing a detector network in the Outer Solar System would massively mitigate the impact of asteroid GGN above $\sim10$\,nHz; we expect that to be quite technically challenging however.
Another available strategy, albeit at possibly prohibitive increased cost and/or technical complexity, could involve flying a multi-constellation Inner Solar System mission with AU baselines, and operating it in the style of the BBO proposal \cite{BBO,Crowder:2005nr} such that correlation of the signals between the different constellations is possible.
We would however expect a noise reduction of only $\sim [N_{\text{constellations}}]^{-1/2}$ in such an approach; see also the discussion in \citeR{PhysRevD.93.021101}.

Another option that is worth investigating further is the use of distant astronomical bodies as test masses. 
Indeed, this option is currently being pursued by terrestrial pulsar timing arrays (of course, terrestrial detectors using \emph{local} proof masses are severely limited below $\sim \text{Hz}$ by terrestrial gravity gradient noise, but this is not a concern when using distant objects as the test masses).
The sensitivity of pulsar timing arrays however decreases above 10\,nHz (see, e.g., \citeR{10.1093/mnras/stv2092}).
To achieve the necessary sensitivity at higher frequencies it may thus be desirable to consider different approaches.

A third tantalizing possibility is offered by astrometry; see, e.g., \citeR[s]{2010IAUS..261..234S,PhysRevLett.119.261102,Wang:2020pmf}. 
A gravitational wave will cause fluctuations in the observed angle between two distant stars. 
By precisely mapping these angular fluctuations, it might be possible to overcome the limitation imposed by asteroid gravity gradient noise in this frequency band. 
In a forthcoming paper, we investigate the technical requirements to realize this possibility to the required levels of strain sensitivity at low frequencies.

\acknowledgments
S.R.~acknowledges travel support from the Gordon and Betty Moore Foundation and the American Physical Society enabling him to visit Stanford to complete this work. 
S.R.~is supported by the U.S.~National Science Foundation (Contract No. PHY-1818899). 
S.R.~is also supported by the DoE under a QuantISED grant for MAGIS.
P.W.G.~and M.F.~would like to express gratitude for the support provided by NSF Grant No.~PHY-1720397, the Heising-Simons Foundation Grant No.~2018-0765, DOE HEP QuantISED Award No.~100495, and the Gordon and Betty Moore Foundation Grant No.~GBMF7946.

\appendix

\section{Technical details}
\label{app:diffAccn}
In this appendix, we give various technical details relevant both to the analytical estimate described in \secref{estimate}, and to the numerical simulation described in \secref{simulation}.

\subsection{Detectors}
\label{app:detectors}

Consider a pair of detectors $I=A,B$ on circular orbits around the Sun at radius $r$, with orbital angular frequency $\Omega = \sqrt{GM_\odot/r^3}$, and separated by a fixed angular separation of  $2\phi$ around the circular orbit, which subtends a fixed baseline distance $L = 2r\sin\phi$.
In a reference system with the origin located at the Sun%
\footnote{\label{ftnt:barycenter}%
    Technically, this should be the Solar System Barycenter, and we should also be using reduced masses throughout; we neglect these technicalities here as they are small, or could be handled reasonably straightforwardly in a refinement of this model.} %
and orientation fixed with respect to the average locations of a distant set of stars, the detectors are located at%
\footnote{\label{ftnt:offset}%
    Note that when written in this way, our numerical simulation actually measures time $t$ from a common temporal offset $t_0 = - \phi / \Omega$  (i.e., we arbitrarily located detector $A$ on the $x$-axis at the start time of the simulation).
    Therefore, the shift $t \rightarrow t - t_0 = t + \phi / \Omega$ should be understood in every equation where relevant throughout the paper (of course, the window function is still taken to taper to zero at the start and end of the simulation).
    Although this temporal offset makes no qualitative difference to our results, we have consistently taken this into account, particularly when considering the signals in \secref{signal} and their comparison to our noise simulations, because it is relevant to the results for simulation No.~3, specifically to the question of the relative sensitivity to the $+$ and $\times$ polarizations given the phasing of the modulation of the detector response.
    Of course, the temporal offset of the start of any possible mission would have an equal-sized impact.
}
\begin{align}
\bm{x}_I(t) &= r \cos\lb( \Omega t \mp \phi \rb) \bm{\hat{x}} + r \sin\lb( \Omega t \mp \phi \rb) \bm{\hat{y}},
\label{eq:detectorCirc}
\end{align}
with the $-$ ($+$) sign for $I=A\ (B)$.
The baseline separation vector between the detectors is
\begin{align}
\bm{r}_{AB}(t) &\equiv \bm{x}_A(t) - \bm{x}_B(t) = L\, \bm{\hat{r}}_{AB}(t);\\
\bm{\hat{r}}_{AB}(t) &\equiv \sin\lb( \Omega t \rb) \bm{\hat{x}} - \cos\lb( \Omega t \rb) \bm{\hat{y}}.
\label{eq:rAB}
\end{align}

\subsection{Differential acceleration for general asteroid/comet orbits}
\label{app:generalOrbits}

Consider an asteroid or comet (`object') $i$ in the JPL-SBD, with mass $M_i$ and with time-dependent location $\bm{X}_i(t)$.

The vectors pointing from each detector $I$ toward object $i$ are $\bm{r}_{i,I}(t) \equiv \bm{X}_i(t) - \bm{x}_I(t)$ and the object--detector separations are $d_{i,I}(t)\equiv | \bm{r}_{i,I}(t) |$.

The vector acceleration of object $i$ on detector $I$ is%
\footnote{\label{ftnt:fixedLocAccn}%
    Note that the accelerations are computed assuming that the positions of the detectors are not disturbed from their circular orbit locations; this is of course an approximation, equivalent to assuming that the asteroid GGN effects are small, as expected.
    } %
\begin{align}
\bm{a}_{i,I}(t) &\equiv GM_i \frac{ \bm{r}_{i,I}(t) }{ \lb[d_{i,I}(t)\rb]^3 }; \quad i = 1,...,N; \quad I = A,B.
\label{eq:accnAstDet}
\end{align}
The total vector acceleration from all \emph{objects} on detector $I$ is
\begin{align}
\bm{a}_{I}(t) &\equiv \sum_{i=1}^{N} \bm{a}_{i,I}(t) ; \quad I = A,B.
\end{align}
Note that $\bm{a}_I$ is \emph{not} the total acceleration on detector $I$, which would of course include the effect of the Sun and planets (and Pluto): that is, the total acceleration on detector $I$ would be $\bm{a}_I^{\text{tot.}} = \bm{a}_I + \sum_{i} \bm{a}_I^{\text{planet }i} + \bm{a}_I^{\text{Sun}}$; we neglect the effect of the planets (and Pluto) on the assumption that they can be removed from the data accurately.

The net relative acceleration on the detector pair from object $i$ projected onto the detector baseline separation vector (`baseline-projected differential acceleration') is
\begin{align}
\Delta a_{i}(t) &\equiv \big[ \bm{a}_{i,A}(t) - \bm{a}_{i,B}(t) \big] \cdot \bm{\hat{r}}_{AB}(t) \label{eq:netproj}
\end{align}
and the total baseline-projected differential acceleration is
\begin{align}
\Delta a(t) &\equiv \big[ \bm{a}_{A}(t)- \bm{a}_{B}(t) \big] \cdot \bm{\hat{r}}_{AB}(t) \label{eq:DeltaADefn}\\
	&=  \lb[\sum_{i=1}^{N}\bm{a}_{i,A}(t) - \sum_{i=1}^{N}\bm{a}_{i,B}(t) \rb] \cdot \bm{\hat{r}}_{AB}(t)
\end{align}
\begin{align}
	&= \sum_{i=1}^{N} \Big\{ \big[ \bm{a}_{i,A}(t) - \bm{a}_{i,B}(t) \big] \cdot \bm{\hat{r}}_{AB}(t) \Big\} \\
	&=  \sum_{i=1}^{N} \Delta a_{i}(t).
\end{align}

We note that we have treated the detector locations as unperturbed from their circular orbital positions [\eqref{detectorCirc}] for the purposes of computing accelerations [\eqref{accnAstDet}] and for computing the detector baseline as a fixed rotating vector [\eqref{rAB}]; we are thus implicitly neglecting second-order-small contributions to $\Delta a_i(t)$.

\subsection{Circular co-planar object orbits}
\label{app:circularCoPlanar}
Consider now the special case of object $i$ with mass $M_i$ on a circular orbit around the Sun (co-planar with the detectors) at radius $R_i$, with orbital angular frequency $\omega_i = \sqrt{GM_\odot/R_i^3}$, and with a phase offset $\alpha_i$ from some fixed zero-reference (with the zero-reference common for all $i$).
Object $i$ is then at the location
\begin{align}
\bm{X}_i(t) &= R_i \cos\lb( \omega_i t + \alpha_i \rb) \bm{\hat{x}} + R_i \sin\lb( \omega_i t + \alpha_i \rb) \bm{\hat{y}}.
\end{align}

In this case, it is possible to make analytical progress in simplifying the  baseline-projected differential acceleration.
The vectors pointing from each detector $I$ toward object $i$ are
\begin{align}
\bm{r}_{i,I}(t)
&= \lb[R_i \cos\lb( \omega_i t + \alpha_i \rb) - r \cos\lb( \Omega t \mp \phi \rb)\rb] \bm{\hat{x}} \nl
    + \lb[ R_i \sin( \omega_i t + \alpha_i ) - r \sin\lb( \Omega t \mp \phi \rb) \rb] \bm{\hat{y}}
\end{align}
where the $-$ ($+$) sign is for $I=A\, (B)$; and the object--detector separations are
\begin{align}
d_{i,I}(t) &= \sqrt{ R_i^2 + r^2 - 2rR_i\cos\lb( \varpi_i t - \alpha_i \mp \phi \rb)}
\end{align}
with the same sign convention, and where $\varpi_i \equiv \Omega-\omega_i$.

The baseline-projected differential acceleration contribution from object $i$ is thus [cf.~\eqref{diffAccn} and footnote \ref{ftnt:schematically}]
\begin{widetext}
\begin{align}
\Delta a_{i}(t)
&= GM_i \Big\{
        R_i \sin(\varpi_i t - \alpha_i) \lb[ d_{i,A}^{-3} - d_{i,B}^{-3} \rb] 
        - r \sin\phi \lb[ d_{i,A}^{-3} + d_{i,B}^{-3}  \rb]  
        \Big\} \\
&= GM_i R_i \sin(\varpi_i t - \alpha_i) \lb\{ \Big[ R_i^2 + r^2 - 2rR_i\cos\lb( \varpi_i t - \alpha_i - \phi \rb)\Big]^{-3/2} -  \Big[ R_i^2 + r^2 - 2rR_i\cos\lb( \varpi_i t - \alpha_i + \phi \rb) \Big]^{-3/2} \rb\} \nl
- GM_i r \sin\phi \lb\{ \Big[ R_i^2 + r^2 - 2rR_i\cos\lb( \varpi_i t - \alpha_i - \phi \rb)\Big]^{-3/2} +  \Big[ R_i^2 + r^2 - 2rR_i\cos\lb( \varpi_i t - \alpha_i + \phi \rb) \Big]^{-3/2} \rb\}.
\label{eq:diffAccnFull}
\end{align}
\end{widetext}

\subsection{Elliptical object orbits}
\label{app:elliptical}
Finally, consider an object $i$ with mass $M_i$ on an elliptical orbit around the Sun with semimajor axis $a_i$, eccentricity $e_i$, argument of perihelion $\hat{\omega}_i$, longitude of the ascending node $\hat{\Omega}_i$, and orbital inclination $\hat{\iota}_i$.%
\footnote{\label{ftnt:hats}%
    We write the orbital element angles with hats to avoid notational conflicts.} %
The angular frequency of the orbit is $\omega_i = \sqrt{GM_\odot/a_i^3}$, and the orbital period is $T_i=2\pi/\omega_i$.
Let the time of perihelion passage be $\tau_{i}$.%
\footnote{\label{ftnt:perihelionPassageTimeChoice}%
    Of course, there is no single time of perihelion passage, so one must select one such time.
    For best accuracy, the specific time chosen should be the epoch of perihelion passage that is nearest in time to the time at which the orbital parameters for a body were determined.
    } %

Let $x\in[0,1]$ be a non-dimensionalized time variable related to true time by $x(t;\tau,a)\equiv (t-\tau)/T(a) \!\!\mod 1$, where $\tau$ and $T(a)$ are, respectively, the time of perihelion passage and the orbital period (the latter of which is fixed by the semimajor axis $a$ by Kepler's Third Law); perihelion passage occurs at $x=0$.
Let $\theta(x;e)$ be the angular position of a body subject to (Newtonian) gravity on an elliptical orbit with eccentricity $e$, evaluated at time $x$.
The function $\theta(x;e)$ must be determined numerically by solving%
\footnote{\label{ftnt:orbitEquation}%
    \eqref{orbitEquation} follows most directly from Kepler's Second Law: $\dot{A}=(1/2)r^2\dot{\theta}=c=\text{const.}$
    Integrating both sides fixes $c=\pi ab/T$ where $a$ and $b=a\sqrt{1-e^2}$ are the semimajor and semiminor axes of the orbit ellipse, respectively.
    Substituting $r=a(1-e^2)/(1+e\cos\theta)$ for an ellipse and changing variables to $x$ completes the derivation.
} %
\begin{align}
    \frac{d\theta(x;e)}{dx} &= \frac{2 \pi}{ \lb( 1 - e^2 \rb)^{ 3/2 } } \Big( 1 + e \cos\big[\theta(x;e)\big] \Big)^2,    \label{eq:orbitEquation}\\
    \theta(0;e) &= 0;
\end{align}
an accurate solution should of course attain $\theta(1,e)=2\pi$ in order for the orbit to be periodic.

Object $i$ is then at the location
\begin{align}
    \bm{X}_i(t) &\equiv r_i(t)\cdot
    R_z(\hat{\Omega}_i) \cdot R_x(\hat{\iota}_i) \cdot R_z(\hat{\omega}_i)
    \cdot \begin{pmatrix} 
            \cos\big[ \theta_i(t) \big] \\[1ex] 
            \sin\big[ \theta_i(t) \big] \\[1ex]
            0
        \end{pmatrix},
    \label{eq:ellipticalOrbit}
\end{align}
where $\theta_i(t)\equiv\theta\big(x(t,\tau_i,a_i);e_i\big)$, $r_i(t) \equiv a_i\lb(1-e_i^2\rb)/\big(1+e_i\cos\big[\theta_i(t)\big]\big)$, and $R_j(\alpha)$ is the $SO(3)$ rotation matrix describing a clockwise rotation around axis $j$ by an angle $\alpha$.
\vspace{-0.1cm}

\section{Multipole expansion of differential acceleration; co-planar, circular orbits}
\label{app:multipole}
Assuming either that $r\ll \min_i R_i$ (detectors inside the asteroid belt), or that $r\gg \max_i R_i$ (detectors outside the asteroid belt), the result \eqref{diffAccnFull} for co-planar circular object and detector orbits is amenable to an analytical multipole expansion.

\subsection{Detectors inside the asteroid belt}
\label{app:multipoleInside}
Take a fixed $i$, and assume that $r\equiv \epsilon R_i$ with $\epsilon\ll 1$.
Additionally, noting that the first $\{\,\cdots\}$-bracket in \eqref{diffAccnFull} vanishes if either $r=0$ or $\phi=0$, and recalling that $L=2r\sin\phi = 2\epsilon R_i\sin\phi$ is the detector baseline, we can write \eqref{diffAccnFull} as
\begin{widetext}
\begin{align}
    \Delta a_i^< \times \lb( \frac{GM_i L }{ 2 R_i^3} \rb)^{-1} &= \frac{\sin(\varpi_i t - \alpha_i)}{\epsilon\sin\phi} \lb\{ \Big[ 1 + \epsilon^2 - 2\epsilon \cos\lb( \varpi_i t - \alpha_i - \phi \rb)\Big]^{-3/2} -  \Big[ 1 + \epsilon^2 - 2\epsilon \cos\lb( \varpi_i t - \alpha_i + \phi \rb) \Big]^{-3/2} \rb\} \nl
- \lb\{ \Big[ 1 + \epsilon^2 - 2\epsilon \cos\lb( \varpi_i t - \alpha_i - \phi \rb)\Big]^{-3/2} +  \Big[ 1 + \epsilon^2 - 2\epsilon\cos\lb( \varpi_i t - \alpha_i + \phi \rb) \Big]^{-3/2} \rb\}.
\label{eq:diffAccnMultipoleInside}
\end{align}
\end{widetext}

It is now straightforward, if tedious, to perform a (multipole)%
\footnote{\label{ftnt:Legendre}%
    Each term  $\sim[\,\cdots]^{-3/2}$ takes the form of the generating function for the derivative of a Legendre polynomial:
         \[ \lb[ 1 + \epsilon^2 - 2\epsilon x\rb]^{-3/2} = \sum_{n=0}^{\infty} P_{n+1}'(x) \epsilon^n.\]
    Writing out the Legendre polynomials explicitly and applying trigonometric identities to reduce powers of trigonometric functions to sums of trigonometric functions of various (higher) frequencies completes the derivation, albeit at the cost of lengthy algebraic manipulations.
    } %
expansion of the two $\{\,\cdots\}$-brackets in powers of $\epsilon$.
While the resulting expressions are algebraically complicated, and we do not show them here in full, it is reasonably straightforward to read off the leading multipole contribution to $\Delta a_i^<$ at each harmonic $\omega_q \equiv q\varpi_i$:
\begin{align}
    \Delta a_i^< &\times \lb( \frac{GM_i L }{ R_i^3} \rb)^{-1} \\
    \supset &\frac{r}{R_i} \frac{3}{4} \cos\phi  \cos\lb( \varpi_i t - \alpha_i \rb) \nl
    	+ \sum_{q=2}^{\infty} \lb( \frac{r}{R_i} \rb)^{q-2} \lb[ -2 \frac{\Gamma(q+1/2)}{\Gamma(1/2)\Gamma(q)} \frac{\sin\lb[(q-1)\phi\rb]}{\sin\phi} \rb] \nl 
	\qquad\qquad\qquad\quad\times \cos\lb[ q\varpi_i t - q\alpha_i \rb],
    \label{eq:inside}
\end{align}
where we have omitted static terms that are present at multipole orders beyond the monopole owing to the non-symmetric distribution of the asteroids outside the detector orbit (the monopole is missing as there is no mass monopole interior to the detector orbit).

In the additional special case where all the masses $M_i=M$ are the same and the asteroids are all at the same radius $R_i=R$ (so that $\varpi_i=\varpi$ are all the same), one can make further progress to understand the total $\Delta a^<$, at least in a statistical sense.
Consider that, in this special case, every leading multipole term in $\Delta a$ at every harmonic contains a term of the form ($q\in \mathbb{Z}$)
\begin{align}
    \Sigma_q &\equiv \sum_{i=1}^N \cos\lb[ q\varpi t - q\alpha_i \rb]
    \equiv X_q \cos\lb( q\varpi t - \xi_q \rb);\\
    X_q &\equiv \lb| \sum_{i=1}^N e^{-iq\alpha_i} \rb|, \quad
    \xi_q \equiv \arg\lb[ \sum_{i=1}^N e^{-iq\alpha_i} \rb].
\end{align}
Assuming that the asteroids are randomly distributed around the circular orbit with uniform probability, such that $\alpha_i \sim \mathcal{U}[0,2\pi)$ where $\mathcal{U}[a,b)$ is the uniform distribution on the half-closed interval $[a,b)$, it follows from the central limit theorem that $\sqrt{2/N} X_q \stackrel{N\ra\infty}{\sim} \chi_2$, where $\chi_\nu$ is the $\chi$ distribution with $\nu$ degrees of freedom.
While it is less relevant, we also have $\xi_q \stackrel{N\ra\infty}{\sim} \mathcal{U}[0,2\pi)$. 
Therefore, for any particular asteroid distribution, the leading multipole terms for each harmonic that appear in $\Delta a^<$ will take the form [cf.~the discussion in \secref{estimate}]
\begin{widetext}
\begin{align}
  \Delta a^<  &\sim c_0 \frac{3\sqrt{\pi}}{8} \lb( \frac{GM L \sqrt{N} }{R^3} \rb)\lb( \frac{r}{R} \rb) \cos\phi  \cos\lb( \varpi t - \xi_0 \rb) \nl
  \quad -\sqrt{\pi}  \lb( \frac{GM L \sqrt{N} }{R^3} \rb) \sum_{q=2}^{\infty} c_q \lb(  \frac{r}{R} \rb)^{q-2}
   \lb[ \frac{\Gamma(q+1/2)}{\Gamma(1/2)\Gamma(q)} \frac{\sin\lb[(q-1)\phi\rb]}{\sin\phi} \rb] \cos\lb[ q\varpi t - \xi_q \rb], \qquad  \text{(special case)}
     \label{eq:specialCaseInside}
\end{align}
\end{widetext}
where $c_q$ are generically $\mathcal{O}(1)$ numbers (respectively, $\xi_q$ are phases) which depend on the exact asteroid distribution.
As the mean of a variable $Z \sim\chi_2$ is $\langle Z \rangle = \sqrt{\pi}/2$, these $c_q$ satisfy $\langle c_q \rangle = 1$, where the ensemble average is taken over random asteroid distributions. 

We have checked \eqref{specialCaseInside} against the computation and analysis pipeline described in \secref{simulation}, with only the minimal changes required to constrain the objects to lie in the requisite equal-radius co-planar circular orbits and all have the same mass; we find excellent agreement at a wide variety of parameter points (we varied the number of objects, the temporal duration of the numerical simulation, the number of discrete points at which the accelerations are computed in the simulation, and the radius of the circular orbit for the objects).
This agreement between the analytical special case and the general numerical pipeline provides strong validation of the results of the latter, and provides increased confidence that the numerical results computed for the general case are correct.

Finally, we make connection with the simplified approximate treatment advanced in \secref{estimate}, at least in the short-baseline limit which is implicitly assumed in that section.
This limit is obtained by sending $\phi \rightarrow 0$ on the RHS of \eqref{diffAccnMultipoleInside} with $L$ held fixed on the LHS.
The result for $\Delta a_i^<$ from \eqref{diffAccnMultipoleInside} in this limit is that given by \eqref{diffAccn}, but with the RHS multiplied by an additional overall factor of
\begin{align}
&-\lb[ 1 -  \frac{3 R_i^2 \sin^2(\varpi t + \alpha_i) }{ R_i^2 + r^2 - 2rR_i \cos( \varpi t + \alpha_i ) } \rb] \\
&\qquad =
 	-\lb[ 1 - \frac{3}{2} \frac{ R_i^2 \lb[ 1 - \cos(2(\varpi t + \alpha_i)) \rb] }{ R_i^2 + r^2 - 2rR_i \cos( \varpi t + \alpha_i ) } \rb],
	\label{eq:extraFactor}
\end{align}
where the $[\,\cdots]$-bracket accounts for geometrical projection and inverse-square force law effects, and the overall sign reflects a difference between the convention choices for the rough arguments advanced in \secref{estimate}, and the specific convention choices in the definition of $\Delta a_i$ at \eqref{DeltaADefn}; as the latter were consistently applied in all the detailed computations in this paper, and the overall sign of the result in \secref{estimate} was not relevant to any of the rough arguments made there, this sign difference is not important.
Because the $\cos(2(\varpi t + \alpha_i))$ term in this correction factor is unsuppressed in an $r/R_i\ll1$ expansion, if it were taken into account in the discussion in \secref{estimate}, it would have been clear that the peak in the Fourier spectrum of the acceleration is at $\omega \sim 2\varpi_i$, that this dominant Fourier mode has the parametric size $GM_i L/R_i^3$, and that the high-frequency tail behaves as $(r/R_i)^{q-2}$; all of which results more closely match those that are clear directly from our more generally applicable result at \eqref{inside}.
However, none of these detailed points modifies any of the qualitative discussion we advanced in \secref{estimate} in a significant way.
We do however note that an unsuppressed correction factor for \eqref{diffAccn} that is $\propto \cos(2(\varpi t + \alpha_i))$ could have been included had we advanced a slightly more careful tidal acceleration argument in \secref{estimate}: geometrical considerations and arguments about the magnitude of $1/r^2$ forces on the proof-masses in various asteroid--baseline orientations, make it clear that $\Delta a_i$ should not only take the form of a tidal acceleration, but also that it should change sign 4 times per complete relative asteroid--detector orbit, with the sign being $+\sigma$ in either fully co-linear orientation of the asteroid and baseline, and $-\sigma$ when the asteroid is located at an angular offset of $\pm \pi/2$ to the baseline direction (here, $\sigma = \pm$ depending on sign conventions chosen to define $\Delta a_i$). 
Clearly, \eqref{diffAccn} does not have this sign-changing property [although we emphasize that \eqref[s]{diffAccnFull} and (\ref{eq:diffAccnMultipoleInside}) both do]; however, multiplying the RHS of \eqref{diffAccn} by $\cos(2(\varpi t + \alpha_i))$ would have given it that property (with the correct phasing), and would have brought the detailed properties of \eqref{diffAccn} more closely in line with those of \eqref{diffAccnMultipoleInside}.
However, since the qualitative features of \eqref{diffAccn} already suffice for the discussion in \secref{estimate}, we elected to omit this correction factor for the sake of simplicity of presentation.

\subsection{Detectors outside the asteroid belt}
\label{app:multipoleOutside}
Take a fixed $i$, and assume that $R_i \equiv \epsilon r$ with $\epsilon\ll 1$.
In this case we can write \eqref{diffAccnFull} as
\begin{widetext}
\begin{align}
\Delta a^>_{i} \times \lb( \frac{GM_iL}{2r^3} \rb)^{-1} 
&=  \epsilon \frac{\sin(\varpi_i t - \alpha_i)}{\sin\phi} \lb\{ \Big[ 1 + \epsilon^2 - 2\epsilon\cos\lb( \varpi_i t - \alpha_i - \phi \rb)\Big]^{-3/2} -  \Big[ 1 + \epsilon^2 - 2\epsilon\cos\lb( \varpi_i t - \alpha_i + \phi \rb) \Big]^{-3/2} \rb\} \nl
- \lb\{ \Big[ 1 + \epsilon^2 - 2\epsilon\cos\lb( \varpi_i t - \alpha_i - \phi \rb)\Big]^{-3/2} +  \Big[ 1 + \epsilon^2 - 2\epsilon\cos\lb( \varpi_i t - \alpha_i + \phi \rb) \Big]^{-3/2} \rb\}.
\label{eq:diffAccnMultipoleOutside}
\end{align}
A similar multipole expansion to that for the `inside the asteroid belt' case yields a slightly different result owing to the rearrangement of the order in $\epsilon$ at which various terms appear in \eqref{diffAccnMultipoleOutside} as compared to \eqref{diffAccnMultipoleInside}; the leading multipole contributions to $\Delta a_i^>$ at each harmonic $\omega_q \equiv q\varpi_i$ are
\begin{align}
\Delta a_{i}^> 
&\supset  \lb( \frac{GM_iL}{r^3} \rb) \sum_{q=1}^{\infty} \lb( \frac{R_i}{r} \rb)^q \lb[ -2\dfrac{ \Gamma\lb[ q+1/2 \rb]}{\Gamma[ 1/2 ]\Gamma[q]} \dfrac{\sin\lb[ ( q - 1) \phi \rb]}{\sin\phi}  - 4\dfrac{ \Gamma\lb[ q + 3/2 \rb]}{\Gamma\lb[1/2 \rb]\Gamma[q+1]}  \cos\lb( q \phi \rb) \rb]\cos\big[ q\varpi_i t - q\alpha_i \big],
\end{align}
where we have dropped a leading unsuppressed static term which is present owing to the existence of a non-zero mass monopole interior to the detector orbits.

Once again, in the additional special case where all the masses $M_i=M$ are the same and the asteroids are all at the same radius $R_i=R$ (so that $\varpi_i=\varpi$ are all the same), a generic random asteroid distribution (with uniform probability for the asteroids to be distributed around their circular orbit at any fixed moment in time) will have
\begin{align}
\Delta a^> 
&\sim - \sqrt{\pi} \lb( \frac{GM L\sqrt{N} }{r^3} \rb) \sum_{q=1}^{\infty} c_q \lb( \frac{R}{r} \rb)^q \lb[\begin{array}{l}
     \dfrac{ \Gamma\lb[ q+1/2 \rb]}{\Gamma[ 1/2 ]\Gamma[q]} \dfrac{\sin\lb[ ( q - 1) \phi \rb]}{\sin\phi}  \\[3ex]
     + 2 \dfrac{ \Gamma\lb[ q + 3/2 \rb]}{\Gamma\lb[1/2 \rb]\Gamma[q+1]}  \cos\lb( q \phi \rb) 
     \end{array} \rb]\cos\big[ q\varpi t - \xi_q \big], \qquad\text{(special case)}
    \label{eq:specialCaseOutside}
\end{align}
\end{widetext}
where the $c_q$ are generically again $\mathcal{O}(1)$ numbers (respectively, $\xi_q$ are phases) that depend on the detailed asteroid distribution and which satisfy the ensemble average $\langle c_q \rangle = 1$.
We have again dropped a static term.

As for the `inside the belt' case, we have also checked \eqref{specialCaseOutside} against the computation and analysis pipeline described in \secref{simulation}, again with only the minimal changes required to constrain the objects to lie in the requisite equal-radius co-planar circular orbits and all have the same mass; we again find excellent agreement at a wide variety of parameter points [we varied the same parameters as discussed in \appref{multipoleInside} below \eqref{specialCaseInside}].
This provides further strong validation of the numerical pipeline for computations involving the general case.

\subsection{Observations on the special case}
\label{app:observations}
An harmonic appearing in \eqref{specialCaseOutside} is two orders higher in the multipole expansion as compared to the same harmonic in \eqref{specialCaseInside}.
Moreover, for the `outside the belt' case in \eqref{specialCaseOutside}, $\varpi \stackrel{r\rightarrow\infty}\rightarrow \Omega$ (i.e., as the detector radius increases larger than the belt radius, the fundamental frequency approaches the asteroid orbital period); by contrast, $\varpi \stackrel{r\rightarrow 0}{\rightarrow} \omega$ for the `inside the belt' case in \eqref{specialCaseInside} (i.e., as the detector orbital radius decreases inside the belt, the fundamental frequency approaches the detector orbital period). 
These observations imply that the noise from asteroids when the detector is located outside the asteroid belt is at lower frequency than the noise when the detector is located inside the asteroid belt.

Also note that if the asteroids were instead exactly periodically distributed around the orbit, which is obviously unphysical, we would obtain instead the deterministic result $X_q = N \delta_{q\!\!\mod N,0}$, where $\delta_{a,b}$ is the Kronecker delta, for both the cases of the detectors inside or outside the belt: there is a parametrically larger acceleration amplitude, by a factor of $\sim\sqrt{N}$, at every $N$-th harmonic as compared to the ensemble average of the randomly distributed case, but the acceleration at all other harmonics is exactly zero owing to phase cancellations.

\section{Discrete Fourier Transform conventions}
\label{app:DFT}
Consider a signal in the time domain $F(t)$ that is sampled at $\mathcal{N}$ discrete points $t_n \equiv n \Delta t$ where $n=0,\hdots,\mathcal{N}-1$ over a duration $T \equiv \mathcal{N} \Delta t$.

Throughout this paper our conventions for the discrete Fourier transform (DFT) of this signal are:
\begin{align}
\tilde{F}(f_k)&= \frac{T}{\mathcal{N}} \sum_{n=0}^{\mathcal{N}-1} F(t_n) e^{ + 2\pi i k n / \mathcal{N} },
\label{eq:DFT}
\end{align}
where $f_k \equiv k \Delta f \equiv k/T$ for $k=0,\hdots,\mathcal{N}-1$, and where we defined the DFT frequency spacing $\Delta f \equiv 1/T$; the inverse DFT (IDFT) is then given by
\begin{align}
F(t_n) &= \frac{1}{T} \sum_{k=0}^{\mathcal{N}-1} \tilde{F}(f_k) e^{-2\pi i k n / \mathcal{N} }.
\label{eq:IDFT}
\end{align}

Under these conventions, the standard multiplication--convolution relationship exists between the time- and frequency-domains, in the form
\begin{align}
C_n &\equiv A_n \cdot B_n \\
\Rightarrow \tilde{C}_k &= \frac{1}{T} \sum_{j=0}^{\mathcal{N}-1} \tilde{A}_j \tilde{B}_{(k-j)\!\!\!\!\mod \mathcal{N}},
\label{eq:convolution}
\end{align}
where we introduced and used the notational shorthand $X_n \equiv X(t_n)$ and $\tilde{X}_k \equiv \tilde{X}(f_k)$.

Both the DFT and IDFT have shift symmetry properties: $F_n = F_{n+\mathcal{N}}$ and $\tilde{F}_{k} = \tilde{F}_{k+\mathcal{N}}$.
Moreover, for a real signal $F(t_n)\in\mathbb{R}$, we have the following DFT reflection symmetry property: $\tilde{F}_k^* = \tilde{F}_{\mathcal{N}-k}$ (real $F_n$).

We define the two-sided power spectral density, $S_k^{(2)}[F] \equiv S^{(2)}(f_k)[F]$, by
\begin{align}
    \langle \lb| F_n \rb|^2 \rangle_T \equiv \sum_{k=0}^{\mathcal{N}-1} \Delta f \times S^{(2)}_k[F] &= \frac{1}{T} \sum_{k=0}^{\mathcal{N}-1} S^{(2)}_k[F], \\
    S^{(2)}_k[F] &= \frac{1}{T} | \tilde{F}_k |^2,
\end{align}
where $\langle X_n \rangle_T \equiv (1/\mathcal{N})\sum_{n=0}^{\mathcal{N}-1} X_n$ is the average value of $X_n$, and where we wrote one factor of $(1/T)$ as $\Delta f$ in the first line to make the Riemann sum nature of this result, which is really just the discrete analog of Parseval's theorem, manifest.

Equivalently, we can define the one-sided power spectral density (PSD), $S_k[F] \equiv S[F](f_k)$, as
\begin{align}
    S_k[F] &\equiv \begin{cases} 
            \dfrac{1}{T} \lb[ |\tilde{F}_k|^2 + |\tilde{F}_{\mathcal{N}-k}|^2 \rb] & k=1,\hdots,\lb\lfloor\tfrac{\mathcal{N}-1}{2}\rb\rfloor\\[2ex]
            \dfrac{1}{T} |\tilde{F}_0|^2 & k=0 \\[2ex]
            \dfrac{1}{T} |\tilde{F}_{\mathcal{N}/2}|^2 & k=\tfrac{\mathcal{N}}{2};\; \mathcal{N}\text{ even} \\
                \end{cases},
        \label{eq:PSD}
\end{align}
in terms of which we have
\begin{align}
    \langle \lb| F_n \rb|^2 \rangle_T &\equiv \sum_{k=0}^{\lb\lceil\tfrac{\mathcal{N}-1}{2}\rb\rceil} \Delta f \times S_k = \frac{1}{T} \sum_{k=0}^{\lb\lceil\tfrac{\mathcal{N}-1}{2}\rb\rceil} S_k.
\end{align}
Note also that for real $F_n$, we have $S_k[F] = \frac{2}{T}|\tilde{F}_k|^2$ for $k=1,\hdots,\lb\lfloor\frac{\mathcal{N}-1}{2}\rb\rfloor$ by the reflection symmetry property discussed above.

Whenever we refer to the PSD without qualification as to whether we mean the one-sided or two-sided version, we implicitly mean the one-sided version.
Note also that the existence of the Nyquist frequency $f_{\text{Nyq.}} \equiv \mathcal{N}/(2T)$ is clear in these results.

For a monochromatic cosine signal with frequency $f = f_q = q\Delta f$ and amplitude $A$, $F(t_n) = A \cos\lb( 2\pi q n / \mathcal{N} \rb)$, we have
\begin{align}
    \tilde{F}_k = T \frac{A}{2} \lb[ \delta_{k,q} + \delta_{k,(\mathcal{N}-q)\!\!\!\!\mod \mathcal{N}} \rb],
    \label{eq:singleCosine}
\end{align}
where $\delta_{a,b}$ is the Kronecker delta; it follows that
\begin{align}
    \tilde{S}_k[F] = T \frac{|A|^2}{2} \delta_{k,q} \lb[ 1 + \delta_{q,0} + \delta_{q,\mathcal{N}/2} \rb].
    \label{eq:monochromaticPSD}
\end{align}

\section{A brief pedagogical introduction to windowing}
\label{app:windowing}
In this appendix we give a brief pedagogical introduction to windowing based on the comprehensive discussion in \citeR{Harris:1978wdg}.

Consider a real, exact cosine signal $s(t) = s_0 \cos(2\pi f t)$ of duration $T$, sampled $\mathcal{N}$ times with cadence $\Delta t$, and with frequency $f = ( n + x ) \Delta f$ where $\Delta f = 1/T$ is the DFT frequency spacing, $n\in \{ 0,\hdots,\lceil (\mathcal{N}-1)/2 \rceil$\} is an integer, and $x\in[0,1)$ is a real number.
There are $f T = n + x$ periods of the signal in the total duration $T$.

Therefore, if $x=0$, there are an exact integer number of periods in the signal duration and it is straightforward to show, as we did at \eqref{singleCosine}, that the DFT of the signal will return Fourier power at exactly one DFT frequency between zero frequency and the Nyquist frequency, inclusive: $f_n = n\Delta f$.

On the other hand, for $x\neq 0$, a non-integer number of periods of the signal is present in the total signal duration; although the DFT will in this case return the most Fourier power at the DFT frequencies nearest the true signal frequency, it is easy to show that Fourier power will appear at \emph{every} DFT frequency between zero frequency and the Nyquist frequency, inclusive.
The phenomenon is known as `spectral leakage'.

\begin{figure}[t]
\includegraphics[width=\columnwidth]{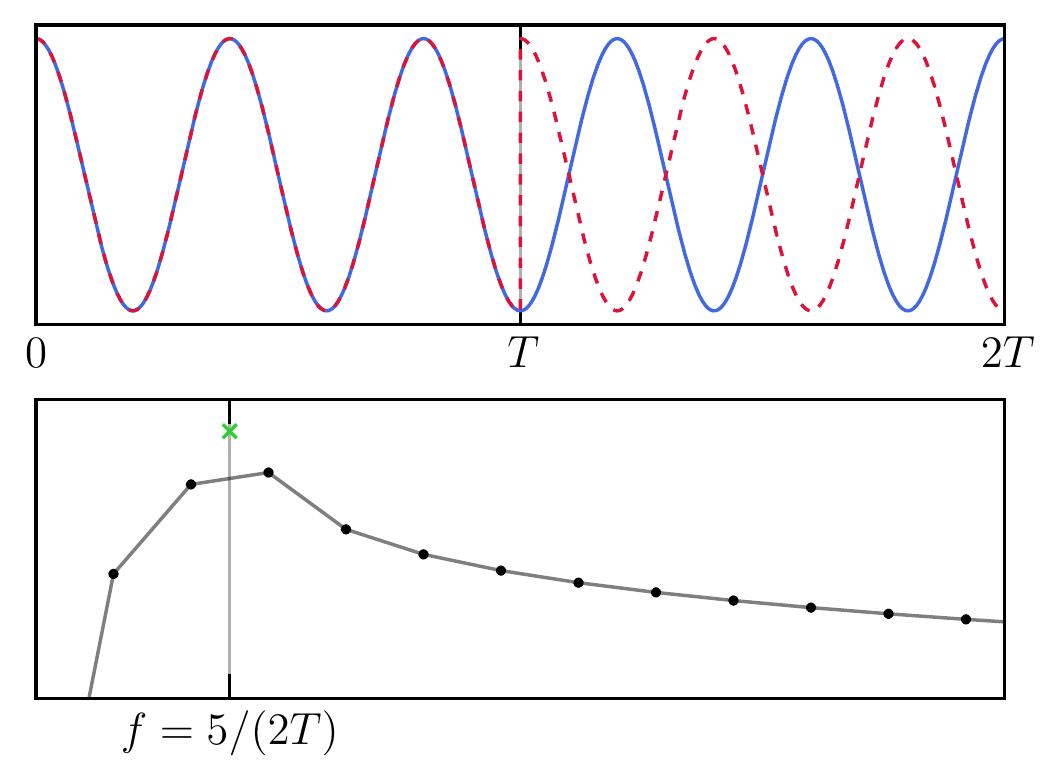}
\caption{\label{fig:discontinuities}%
    \textsc{Top panel:}
    An example of the discontinuity in the time-domain that appears in an IDFT-reconstructed signal when the original signal (solid blue)---here a cosine with frequency $f = 5/(2T)$ defined on the interval $[0,2T$]---is analyzed by DFT over a duration---here, $[0,T]$---that does not contain an integer number of full periods, and is then reconstructed into the time domain by IDFT (dashed red) and examined outside the interval $[0,T]$ (formally, by allowing $n$ in \eqref{IDFT} to run over integers outside the range $n=0,\hdots,\mathcal{N}-1$).
    It is clear that the original and reconstructed signals agree on the interval $[0,T]$ over which the DFT was performed, but that they disagree on the next duration-$T$ interval.
    \textsc{Bottom panel:}
    Displayed on a log-linear scale in the frequency domain, the one-sided PSD of the signal considered over the temporal duration $[0,T]$ (black dots; grey solid line joins points to guide the eye) clearly shows spectral leakage.
    This leakage can be thought of as arising from the discontinuity in the reconstructed signal at time $T$.
    For comparison, we also show the PSD of the original signal considered over the temporal duration $[0,2T]$ (green cross; values at all other frequencies are exactly zero), on which it is exactly periodic; this does not display any spectral leakage, as expected.
	} 
\end{figure}

Spectral leakage can be understood intuitively in the following way.
Consider a signal $s(t)$ in the time domain (and assume the signal is defined on a time period longer than $T$) that is analyzed into the frequency domain by DFT as $\tilde{s}(f)$ using only a duration $T$ of time-domain data.
Now reconstruct the signal into the time domain from only those frequency-domain data via the IDFT, as $\hat{s}(t)$.
Because of the IDFT discrete shift symmetry discussed in \appref{DFT}, when $\hat{s}(t)$ is considered on durations longer than $T$, it will automatically exhibit a periodicity with period $T$.
By contrast, a signal that does not have an exact integer number of periods within a duration $T$ cannot have periodicity with period $T$: i.e., $s(t_0) \neq s(T_\mathcal{N})$.
This mismatch between the imposed periodicity of the IDFT, $\hat{s}(t_0) = \hat{s}(t_\mathcal{N})$, and the true signal behavior, $s(t_0) \neq s(T_\mathcal{N})$, implies that the IDFT-reconstructed signal is forced to exhibit discontinuities when viewed on timescales longer than $T$; see \figref{discontinuities} for an example.
Intuitively, this leads to spectral leakage because the Fourier transform of a discontinuity exhibits power at every frequency.

\begin{figure*}[p]
\includegraphics[width=1.4\columnwidth]{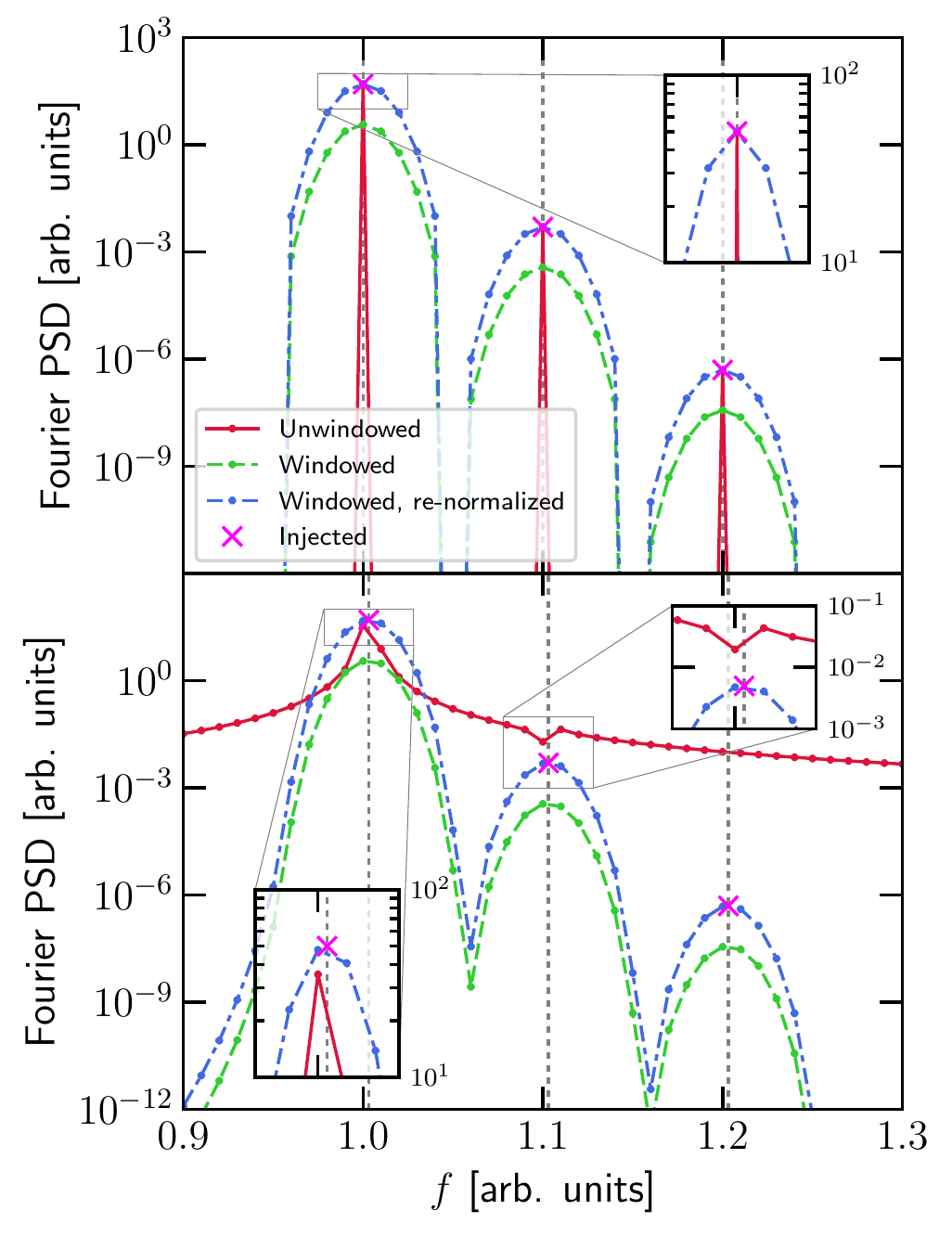}
\caption{\label{fig:windowing}%
    In both panels, the red line shows the one-sided Fourier PSD of an unwindowed signal; the green dashed line shows the PSD of the windowed signal (see text); the blue dash-dot line shows the PSD of the windowed signal renormalized by a factor of $(a_0)^{-2}$ [see text] to obtain the correct expected peak PSD values at the injected frequencies; and the fuchsia crosses show the expected PSD values for the true signal at the true frequencies given the true signal amplitudes and the signal duration $T$.
    Note that the DFT, and hence PSD, is only defined at discrete frequency values $f_k = k \Delta f = k/T$ for $k=0,\hdots,\mathcal{N}-1$ (marked with solid dots on all lines); for clarity, we have joined these points with straight lines in these plots.
    	\textsc{Top panel:} %
    	    The signal is a sum of three exact cosine terms with vastly different amplitudes, with the frequencies of the three terms all being exact multiples of $\Delta f$. 
    	    As indicated by vertical gray dotted lines, the spacing between successive frequency components is $10 \times \Delta f$.
    	    Here, the window function shows no advantages: it smears out the Fourier power from the expected frequencies to neighboring frequencies, lowering peak power and decreasing resolving power for signals of similar amplitude that lie within the width of the Fourier representation of the window function.
    	    The re-normalized windowed PSD recovers the expected signal power at the expected signal frequencies; see inset for detail.
       	\textsc{Bottom panel:} %
       	    The same signal as in the top panel, with the exception that all three of the signal components have been shifted up in frequency by a (common) irrational fraction of the DFT frequency spacing $\Delta f$ so that they all lie between successive DFT frequencies, as shown by the vertical gray dotted lines.
        	Here, the spectral leakage from the dominant term largely masks the presence of the other two signal components in the unwindowed PSD. 
        	A feature corresponding to the middle signal component is noticeable (see upper inset); it is however distorted, as the signal power from that component is similar in magnitude to the leakage power from the dominant component in the vicinity of the frequency of the middle component.
        	On the other hand, the windowed result clearly reveals the presence of all three signal components without any of these issues, albeit at the expense of smearing out the Fourier power from the expected frequencies to neighboring frequencies, and reducing peak power.
        	This resolution--dynamic-range trade-off is a generic feature of windowing, as discussed in the text.
            As shown in the lower inset axes, the re-normalized windowed PSD again recovers quite well the expected signal power, and importantly does a better job at locating the expected signal frequencies as compared to the unwindowed result; cf.~the unwindowed PSD peak for the the highest-power component is shifted from the injected frequency.
	} 
\end{figure*}

The salient issue that arises as a result of the spectral leakage phenomenon is that low-power generic (i.e., not at an exact DFT frequency) signals that appear in the vicinity (in the frequency domain) of high-power generic signals can get swamped by the spectral leakage from the higher-power signal, making their extraction difficult; see lower panel of \figref{windowing}.
The classical resolution of this problem is to `window' (or `apodize', or `taper') the data: that is, to multiply the original time-domain data by a continuous (and usually smooth) function $w(t)$ that is symmetric $w(t) = w(T-t)$, satisfies $w(T/2) = 1$, and typically satisfies $w(0)=w(T)=0$ (or, at the very least, decays significantly at the end points: $w(0)=w(T)\ll w(T/2)=1$), before taking the DFT: $s(t) \rightarrow s'(t) \equiv s(t) w(t)$.
While there are an infinity of possible choices for the exact form for $w(t)$, the goal in all cases is the same: to modify the nature of the spectral leakage by de-emphasizing the importance of the edges of the signal duration, so as to minimize the discontinuity effects discussed above that intuitively give rise to the leakage.

Windowing is a multiplication in the time domain, so it is a convolution in the frequency domain; cf.~\eqref{convolution}.
This observation allows one to understand the fundamental trade-off that arises when choosing a window function: dynamic range vs.~resolution.
Suppose the goal is to minimize spectral leakage as much as possible: intuitively, one would guess (correctly) that the appropriate window function to choose would be one that approaches $w(t)\rightarrow 0$ at the endpoints of the duration-$T$ interval as smoothly as possible, in order to suppress any of the discontinuity effects discussed above as the intuitive reason for the leakage.
However, a time-domain window function that approaches the end points very smoothly will, when viewed in the frequency domain, necessarily contain more frequency components than a time-domain window function that, e.g., simply goes to zero but with a discontinuous first derivative at the endpoints.
That is, a more smoothly tapered window function will be broader and have lower peak power in the frequency domain than a less smoothly tapered window function.
Because the frequency-domain representation of the window function gets convolved with the frequency-domain representation of the true signal to obtain the frequency-domain representation of the windowed monochromatic signal, the latter will be more smeared out and have lower peak power when a smoother window is used as compared to a less smooth window.
Therefore, improved ability to resolve signals of more disparate power that are not very closely spaced in frequency (i.e., higher dynamic range) comes at the cost of impaired ability to resolve signals of similar power that are closely spaced in frequency (i.e., worse resolution), and vice versa.
These effects are clearly evident in \figref{windowing}.

In this paper, we utilize throughout a simple window function consisting of a sum of five exact cosine terms that respect the reflection symmetry $w(t) = w(T-t)$, with coefficients chosen to zero the window at the endpoints, $w(0)=0$; zero as many of the derivatives as possible at the endpoints, $w^{(n)}(0)=0$ for $n=1,\hdots,6$ (some identically, as there would otherwise be insufficiently many free parameters in the definition below); and maintain the normalization $w(T/2)=1$:
\begin{align}
    w(t) &= \sum_{n=0}^{4} a_n \cos\lb[2\pi n \lb(\frac{t}{T}-\frac{1}{2}\rb) \rb]
\end{align}
with
\begin{align}
    \{a_0,a_1,a_2,a_3,a_4\} &= \lb\{ \frac{35}{128}, \frac{7}{16}, \frac{7}{32}, \frac{1}{16}, \frac{1}{128} \rb\}
\end{align}
This simplifies to 
\begin{align}
   w(t) &= \sin^8\lb( \pi t/T \rb).
   \label{eq:window}
\end{align}

Because windowing is a convolution in the frequency domain, the peak DFT amplitude of a windowed monochromatic signal that lies at an exact DFT frequency would simply be the DFT amplitude of the unwindowed signal multiplied by the coefficient of the zero-frequency component of the window function; i.e., a factor of $a_0 \approx 0.28$ in the DFT, or a factor of $(a_0)^2\approx 7.5\times 10^{-2}$ in the one-sided PSD.
To recover the correct peak power, we could thus re-normalize the PSD of the windowed result by $(a_0)^{-2} \approx 13$; see the blue dot-dashed lines in \figref{windowing}.
Also note that we have $\int_0^T w(t) dt = a_0 =35/128\approx 0.28$, and\linebreak $\frac{1}{T} \int_0^T [w(t)]^2 =  a_0 + (1/2) \sum_{i=1}^4 a_i^2 = 6435/32768 \approx 0.2$.

\bibliographystyle{JHEP}
\bibliography{references.bib}

\providecommand{\href}[2]{#2}\begingroup\justify\begin{thebibliography}{10}

\bibitem{PhysRevLett.116.061102}
{\scshape LIGO Scientific Collaboration and Virgo Collaboration}, B.~P. Abbott,
  R.~Abbott, T.~D. Abbott, M.~R. Abernathy, F.~Acernese, K.~Ackley et~al.,
  \emph{Observation of gravitational waves from a binary black hole merger},
  \href{http://dx.doi.org/10.1103/PhysRevLett.116.061102}{\emph{Phys. Rev.
  Lett.} {\bf 116} (2016) 061102}.

\bibitem{PhysRevLett.119.161101}
{\scshape LIGO Scientific Collaboration and Virgo Collaboration}, B.~P. Abbott,
  R.~Abbott, T.~D. Abbott, F.~Acernese, K.~Ackley, C.~Adams et~al.,
  \emph{{GW170817: Observation of Gravitational Waves from a Binary Neutron
  Star Inspiral}},
  \href{http://dx.doi.org/10.1103/PhysRevLett.119.161101}{\emph{Phys. Rev.
  Lett.} {\bf 119} (2017) 161101}.

\bibitem{PhysRevX.9.031040}
{\scshape LIGO Scientific Collaboration and Virgo Collaboration}, B.~P. Abbott,
  R.~Abbott, T.~D. Abbott, S.~Abraham, F.~Acernese, K.~Ackley et~al.,
  \emph{{GWTC-1: A Gravitational-Wave Transient Catalog of Compact Binary
  Mergers Observed by LIGO and Virgo during the First and Second Observing
  Runs}}, \href{http://dx.doi.org/10.1103/PhysRevX.9.031040}{\emph{Phys. Rev.
  X} {\bf 9} (2019) 031040}.

\bibitem{Abbott:2020niy}
{\scshape LIGO Scientific Collaboration and Virgo Collaboration}, R.~Abbott
  et~al., \emph{{GWTC-2: Compact Binary Coalescences Observed by LIGO and Virgo
  During the First Half of the Third Observing Run}},
  \href{https://arxiv.org/abs/2010.14527}{{\tt arXiv:2010.14527}}.

\bibitem{Akutsu:2020zlw}
{\scshape KAGRA Collaboration}, T.~Akutsu et~al., \emph{{Overview of KAGRA:
  KAGRA science}},  \href{https://arxiv.org/abs/2008.02921}{{\tt
  arXiv:2008.02921}}.

\bibitem{Sesana:2019vho}
A.~Sesana et~al., \emph{{Unveiling the Gravitational Universe at $\mu$-Hz
  Frequencies}},  \href{https://arxiv.org/abs/1908.11391}{{\tt
  arXiv:1908.11391}}.

\bibitem{Baibhav:2019rsa}
V.~Baibhav et~al., \emph{{Probing the Nature of Black Holes: Deep in the mHz
  Gravitational-Wave Sky}},  \href{https://arxiv.org/abs/1908.11390}{{\tt
  arXiv:1908.11390}}.

\bibitem{Sedda:2019uro}
M.~A. Sedda et~al., \emph{{The missing link in gravitational-wave astronomy:
  discoveries waiting in the decihertz range}},
  \href{http://dx.doi.org/10.1088/1361-6382/abb5c1}{\emph{Class. Quant. Grav.}
  {\bf 37} (2020) 215011} [\href{https://arxiv.org/abs/1908.11375}{{\tt
  arXiv:1908.11375}}].

\bibitem{Baker:2019pnp}
J.~Baker et~al., \emph{{Space Based Gravitational Wave Astronomy Beyond LISA}},
   \href{https://arxiv.org/abs/1907.11305}{{\tt arXiv:1907.11305}}.

\bibitem{Kramer_2013}
M.~Kramer and D.~J. Champion, \emph{{The European Pulsar Timing Array and the
  Large European Array for Pulsars}},
  \href{http://dx.doi.org/10.1088/0264-9381/30/22/224009}{\emph{Class. Quant.
  Grav.} {\bf 30} (2013) 224009}.

\bibitem{Kerr_2020}
M.~Kerr, D.~J. Reardon, G.~Hobbs, R.~M. Shannon, R.~N. Manchester, S.~Dai
  et~al., \emph{{The Parkes Pulsar Timing Array project: second data release}},
  \href{http://dx.doi.org/10.1017/pasa.2020.11}{\emph{Publ. Astron. Soc. Aust.}
  {\bf 37} (2020) e020}.

\bibitem{Arzoumanian:2020vkk}
{\scshape NANOGrav Collaboration}, Z.~Arzoumanian et~al., \emph{{The NANOGrav
  12.5 yr Data Set: Search for an Isotropic Stochastic Gravitational-wave
  Background}},
  \href{https://iopscience.iop.org/article/10.3847/2041-8213/abd401}{\emph{Astrophys.
  J. Lett.} {\bf 905} (2020) L34} [\href{https://arxiv.org/abs/2009.04496}{{\tt
  arXiv:2009.04496}}].

\bibitem{2016MNRAS.458.1267V}
J.~P.~W. {Verbiest}, L.~{Lentati}, G.~{Hobbs}, R.~{van Haasteren}, P.~B.
  {Demorest}, G.~H. {Janssen} et~al., \emph{{The International Pulsar Timing
  Array: First data release}},
  \href{http://dx.doi.org/10.1093/mnras/stw347}{\emph{Mon. Not. R. Astron.
  Soc.} {\bf 458} (2016) 1267--1288}
  [\href{https://arxiv.org/abs/1602.03640}{{\tt arXiv:1602.03640}}].

\bibitem{Baker:2019nia}
J.~Baker et~al., \emph{{The Laser Interferometer Space Antenna: Unveiling the
  Millihertz Gravitational Wave Sky}},
  \href{https://arxiv.org/abs/1907.06482}{{\tt arXiv:1907.06482}}.

\bibitem{LISA_Sci_Req}
{\scshape LISA Science Study Team},
  \emph{\href{https://www.cosmos.esa.int/documents/678316/1700384/SciRD.pdf}{LISA
  Science Requirements Document}},  Tech. Rep. ESA-L3-EST-SCI-RS-001, 2018.

\bibitem{amaroseoane2017laser}
P.~Amaro-Seoane, H.~Audley, S.~Babak, J.~Baker, E.~Barausse, P.~Bender et~al.,
  \emph{{Laser Interferometer Space Antenna}},
  \href{https://arxiv.org/abs/1702.00786}{{\tt arXiv:1702.00786}}.

\bibitem{PhysRevLett.120.061101}
M.~Armano, H.~Audley, J.~Baird, P.~Binetruy, M.~Born, D.~Bortoluzzi et~al.,
  \emph{{Beyond the Required LISA Free-Fall Performance: New LISA Pathfinder
  Results down to $20\text{ }\text{ }\ensuremath{\mu}\mathrm{Hz}$}},
  \href{http://dx.doi.org/10.1103/PhysRevLett.120.061101}{\emph{Phys. Rev.
  Lett.} {\bf 120} (2018) 061101}.

\bibitem{Luo:2015ght}
J.~Luo et~al., \emph{{TianQin: a space-borne gravitational wave detector}},
  \href{http://dx.doi.org/10.1088/0264-9381/33/3/035010}{\emph{Class. Quant.
  Grav.} {\bf 33} (2016) 035010} [\href{https://arxiv.org/abs/1512.02076}{{\tt
  arXiv:1512.02076}}].

\bibitem{Milyukov:2020fm}
V.~K. Milyukov, \emph{{TianQin Space-Based Gravitational Wave Detector: Key
  Technologies and Current State of Implementation}},
  \href{http://dx.doi.org/10.1134/S1063772920120070}{\emph{Astronomy Reports}
  {\bf 64} (2020) 1067--1077}.

\bibitem{Dimopoulos:2007cj}
S.~Dimopoulos, P.~W. Graham, J.~M. Hogan, M.~A. Kasevich and S.~Rajendran,
  \emph{{Gravitational Wave Detection with Atom Interferometry}},
  \href{http://dx.doi.org/10.1016/j.physletb.2009.06.011}{\emph{Phys. Lett. B}
  {\bf 678} (2009) 37--40} [\href{https://arxiv.org/abs/0712.1250}{{\tt
  arXiv:0712.1250}}].

\bibitem{Dimopoulos:2008sv}
S.~Dimopoulos, P.~W. Graham, J.~M. Hogan, M.~A. Kasevich and S.~Rajendran,
  \emph{{An Atomic Gravitational Wave Interferometric Sensor (AGIS)}},
  \href{http://dx.doi.org/10.1103/PhysRevD.78.122002}{\emph{Phys. Rev. D} {\bf
  78} (2008) 122002} [\href{https://arxiv.org/abs/0806.2125}{{\tt
  arXiv:0806.2125}}].

\bibitem{Hogan:2010fz}
J.~M. Hogan et~al., \emph{{An Atomic Gravitational Wave Interferometric Sensor
  in Low Earth Orbit (AGIS-LEO)}},
  \href{http://dx.doi.org/10.1007/s10714-011-1182-x}{\emph{Gen. Rel. Grav.}
  {\bf 43} (2011) 1953--2009} [\href{https://arxiv.org/abs/1009.2702}{{\tt
  arXiv:1009.2702}}].

\bibitem{Graham:2017pmn}
{\scshape MAGIS Collaboration}, P.~W. Graham, J.~M. Hogan, M.~A. Kasevich,
  S.~Rajendran and R.~W. Romani, \emph{{Mid-band gravitational wave detection
  with precision atomic sensors}},
  \href{https://arxiv.org/abs/1711.02225}{{\tt arXiv:1711.02225}}.

\bibitem{Coleman:2018ozp}
{\scshape MAGIS-100 Collaboration}, J.~Coleman, \emph{{Matter-wave Atomic
  Gradiometer Interferometric Sensor (MAGIS-100) at Fermilab}},
  \href{http://dx.doi.org/10.22323/1.340.0021}{\emph{PoS} {\bf ICHEP2018}
  (2019) 021} [\href{https://arxiv.org/abs/1812.00482}{{\tt
  arXiv:1812.00482}}].

\bibitem{Canuel:2018fq}
B.~Canuel, A.~Bertoldi, L.~Amand, E.~Pozzo~di Borgo, T.~Chantrait, C.~Danquigny
  et~al., \emph{Exploring gravity with the {MIGA} large scale atom
  interferometer},
  \href{http://dx.doi.org/10.1038/s41598-018-32165-z}{\emph{Sci. Rep.} {\bf 8}
  (2018) 14064}.

\bibitem{Badurina:2019hst}
L.~Badurina et~al., \emph{{AION: An Atom Interferometer Observatory and
  Network}}, \href{http://dx.doi.org/10.1088/1475-7516/2020/05/011}{\emph{JCAP}
  {\bf 05} (2020) 011} [\href{https://arxiv.org/abs/1911.11755}{{\tt
  arXiv:1911.11755}}].

\bibitem{Tino:2019tkb}
G.~M. Tino et~al., \emph{{SAGE: A Proposal for a Space Atomic Gravity
  Explorer}}, \href{http://dx.doi.org/10.1140/epjd/e2019-100324-6}{\emph{Eur.
  Phys. J. D} {\bf 73} (2019) 228}
  [\href{https://arxiv.org/abs/1907.03867}{{\tt arXiv:1907.03867}}].

\bibitem{BBO}
E.~S. Phinney et~al., \emph{{The Big Bang Observer: direct detection of
  gravitational waves from the birth of the universe to the present}},  NASA
  Mission Concept Study, 2004.

\bibitem{Crowder:2005nr}
J.~Crowder and N.~J. Cornish, \emph{{Beyond LISA: Exploring future
  gravitational wave missions}},
  \href{http://dx.doi.org/10.1103/PhysRevD.72.083005}{\emph{Phys. Rev. D} {\bf
  72} (2005) 083005} [\href{https://arxiv.org/abs/gr-qc/0506015}{{\tt
  arXiv:gr-qc/0506015}}].

\bibitem{Seto:2001qf}
N.~Seto, S.~Kawamura and T.~Nakamura, \emph{{Possibility of direct measurement
  of the acceleration of the universe using 0.1-Hz band laser interferometer
  gravitational wave antenna in space}},
  \href{http://dx.doi.org/10.1103/PhysRevLett.87.221103}{\emph{Phys. Rev.
  Lett.} {\bf 87} (2001) 221103}
  [\href{https://arxiv.org/abs/astro-ph/0108011}{{\tt
  arXiv:astro-ph/0108011}}].

\bibitem{Kawamura:2018esd}
S.~Kawamura et~al., \emph{{Space gravitational-wave antennas DECIGO and
  B-DECIGO}}, \href{http://dx.doi.org/10.1142/S0218271818450013}{\emph{Int. J.
  Mod. Phys. D} {\bf 28} (2019) 1845001}.

\bibitem{kawamura2020current}
S.~Kawamura, M.~Ando, N.~Seto, S.~Sato, M.~Musha, I.~Kawano et~al.,
  \emph{{Current status of space gravitational wave antenna DECIGO and
  B-DECIGO}},  \href{https://arxiv.org/abs/2006.13545}{{\tt arXiv:2006.13545}}.

\bibitem{Maggiore:2019uih}
M.~Maggiore et~al., \emph{{Science Case for the Einstein Telescope}},
  \href{http://dx.doi.org/10.1088/1475-7516/2020/03/050}{\emph{JCAP} {\bf 03}
  (2020) 050} [\href{https://arxiv.org/abs/1912.02622}{{\tt
  arXiv:1912.02622}}].

\bibitem{Harms:2015awd}
J.~Harms, \emph{Terrestrial gravity fluctuations},
  \href{http://dx.doi.org/10.1007/lrr-2015-3}{\emph{Living Rev. Relativ.} {\bf
  18} (2015) 3}.

\bibitem{PhysRevD.86.102001}
J.~C. Driggers, J.~Harms and R.~X. Adhikari, \emph{Subtraction of {N}ewtonian
  noise using optimized sensor arrays},
  \href{http://dx.doi.org/10.1103/PhysRevD.86.102001}{\emph{Phys. Rev. D} {\bf
  86} (2012) 102001}.

\bibitem{Coughlin_2016}
M.~Coughlin, N.~Mukund, J.~Harms, J.~Driggers, R.~Adhikari and S.~Mitra,
  \emph{{Towards a first design of a Newtonian-noise cancellation system for
  Advanced {LIGO}}},
  \href{http://dx.doi.org/10.1088/0264-9381/33/24/244001}{\emph{Class. Quant.
  Grav.} {\bf 33} (2016) 244001}.

\bibitem{PhysRevLett.121.221104}
M.~W. Coughlin, J.~Harms, J.~Driggers, D.~J. McManus, N.~Mukund, M.~P. Ross
  et~al., \emph{{Implications of Dedicated Seismometer Measurements on
  Newtonian-Noise Cancellation for Advanced LIGO}},
  \href{http://dx.doi.org/10.1103/PhysRevLett.121.221104}{\emph{Phys. Rev.
  Lett.} {\bf 121} (2018) 221104}.

\bibitem{Buikema_2020}
A.~Buikema, C.~Cahillane, G.~Mansell, C.~Blair, R.~Abbott, C.~Adams et~al.,
  \emph{{Sensitivity and performance of the Advanced LIGO detectors in the
  third observing run}},
  \href{http://dx.doi.org/10.1103/physrevd.102.062003}{\emph{Phys. Rev. D} {\bf
  102} (2020) 062003}.

\bibitem{LISA-Pre-Phase-A}
{\scshape LISA Study Team},
  \emph{\href{https://lisa.nasa.gov/archive2011/Documentation/ppa2.08.pdf}{LISA
  Pre-Phase A Report}},  Tech. Rep. MPQ 233 (second ed.), 1998.

\bibitem{JPL-SBD}
\url{https://ssd.jpl.nasa.gov} (accessed 2020).

\bibitem{JPL-SBD-phys}
\url{https://ssd.jpl.nasa.gov/?phys_data} (accessed 2020).

\bibitem{2014ApJ...791..121M}
J.~R. {Masiero}, T.~{Grav}, A.~K. {Mainzer}, C.~R. {Nugent}, J.~M. {Bauer},
  R.~{Stevenson} et~al., \emph{{Main-belt Asteroids with WISE/NEOWISE:
  Near-infrared Albedos}},
  \href{http://dx.doi.org/10.1088/0004-637X/791/2/121}{\emph{\apj} {\bf 791}
  (2014) 121} [\href{https://arxiv.org/abs/1406.6645}{{\tt arXiv:1406.6645}}].

\bibitem{JPL-HORIZONS}
\url{https://ssd.jpl.nasa.gov/?horizons} (accessed 2020).

\bibitem{FOWLER_Asteroids}
J.~W. Fowler and J.~Chillemi, \emph{{IRAS} asteroid data processing},  in
  \emph{{The IRAS Minor Planet Survey}}, Tech. Rep. PL-TR-92--2049, pp.~17--43.
\newblock Philips Laboratory, Hanscom AF Base, MA, 1992.

\bibitem{HARRIS1997450}
A.~W. Harris and A.~W. Harris, \emph{On the revision of radiometric albedos and
  diameters of asteroids},
  \href{http://dx.doi.org/https://doi.org/10.1006/icar.1996.5664}{\emph{Icarus}
  {\bf 126} (1997) 450--454}.

\bibitem{Harris:1978wdg}
F.~J. Harris, \emph{{On the Use of Windows for Harmonic Analysis with the
  Discrete Fourier Transform}},
  \href{http://dx.doi.org/10.1109/PROC.1978.10837}{\emph{{Proc. IEEE}} {\bf 66}
  (1978) 51--83}.

\bibitem{Misner:1974qy}
C.~W. Misner, K.~Thorne and J.~Wheeler, \emph{{Gravitation}}.
\newblock W. H. Freeman, San Francisco, 1973.

\bibitem{Maggiore:2007zz}
M.~Maggiore, \emph{{Gravitational Waves. Volume 1: Theory and Experiments}}.
\newblock Oxford University Press, Oxford, 2008.

\bibitem{Moore:2014lga}
C.~Moore, R.~Cole and C.~Berry, \emph{{Gravitational-wave sensitivity curves}},
  \href{http://dx.doi.org/10.1088/0264-9381/32/1/015014}{\emph{Class. Quant.
  Grav.} {\bf 32} (2015) 015014} [\href{https://arxiv.org/abs/1408.0740}{{\tt
  arXiv:1408.0740}}].

\bibitem{Shoemaker:1983qae}
E.~M. Shoemaker, \emph{{Asteroid and Comet Bombardment of the Earth}},
  \href{http://dx.doi.org/10.1146/annurev.ea.11.050183.002333}{\emph{Ann. Rev.
  Earth Planet. Sci} {\bf 11} (1983) 461--494}.

\bibitem{PhysRevD.93.021101}
W.~Chaibi, R.~Geiger, B.~Canuel, A.~Bertoldi, A.~Landragin and P.~Bouyer,
  \emph{Low frequency gravitational wave detection with ground-based atom
  interferometer arrays},
  \href{http://dx.doi.org/10.1103/PhysRevD.93.021101}{\emph{Phys. Rev. D} {\bf
  93} (2016) 021101} [\href{https://arxiv.org/abs/1601.00417}{{\tt
  arXiv:1601.00417}}].

\bibitem{10.1093/mnras/stv2092}
S.~Babak, A.~Petiteau, A.~Sesana, P.~Brem, P.~A. Rosado, S.~R. Taylor et~al.,
  \emph{{European Pulsar Timing Array limits on continuous gravitational waves
  from individual supermassive slack hole binaries}},
  \href{http://dx.doi.org/10.1093/mnras/stv2092}{\emph{Mon. Not. R. Astron.
  Soc.} {\bf 455} (2016) 1665--1679}.

\bibitem{2010IAUS..261..234S}
B.~F. {Schutz},
  \emph{{\href{https://doi.org/10.1017/S1743921309990457}{Astrometric and
  timing effects of gravitational waves}}},  in \emph{Relativity in Fundamental
  Astronomy: Dynamics, Reference Frames, and Data Analysis} (S.~A. {Klioner},
  P.~K. {Seidelmann} and M.~H. {Soffel}, eds.), vol.~261, pp.~234--239, 2010.

\bibitem{PhysRevLett.119.261102}
C.~J. Moore, D.~P. Mihaylov, A.~Lasenby and G.~Gilmore, \emph{{Astrometric
  Search Method for Individually Resolvable Gravitational Wave Sources with
  Gaia}}, \href{http://dx.doi.org/10.1103/PhysRevLett.119.261102}{\emph{Phys.
  Rev. Lett.} {\bf 119} (2017) 261102}.

\bibitem{Wang:2020pmf}
Y.~Wang, K.~Pardo, T.-C. Chang and O.~Dor{\'e}, \emph{Gravitational wave
  detection with photometric surveys},
  \href{http://dx.doi.org/10.1103/physrevd.103.084007}{\emph{Phys. Rev. D} {\bf
  103} (2021) 084007} [\href{https://arxiv.org/abs/2010.02218}{{\tt
  arXiv:2010.02218}}].

\end{thebibliography}\endgroup

\end{document}